\documentclass[a4paper,fleqn,usenatbib]{mnras}
\usepackage[T1]{fontenc}
\usepackage{ae,aecompl}
\usepackage{amstext}
\usepackage{url}
\usepackage{graphicx,epsfig,color,latexsym,hyperref}	
\usepackage{tabularx,natbib,bm,times}
\usepackage{subfigure}
\usepackage{amsmath}	
\usepackage{amssymb}	
\usepackage{multirow}

\def\elabel#1{\label{eq:#1}}

\newcommand{\bea}{\begin{eqnarray}}
\newcommand{\eea}{\end{eqnarray}}
\newcommand{\be}{\begin{equation}}
\newcommand{\ee}{\end{equation}}

\title[Deep Learning for Strong Lensing Search]{Deep Learning for Strong Lensing Search: Tests of the Convolutional Neural Networks and New Candidates from KiDS DR3}
\author[Z. He et al.]
	{Zizhao He$^1$\thanks{E-mail: leaveformoon@hotmail.com}, Xinzhong Er$^1$\thanks{E-mail: phioen@163.com}, Qian Long$^2$\thanks{E-mail: longqian@ynao.ac.cn}, Dezi Liu$^{1,3}$, Xiangkun Liu$^1$, Ziwei Li$^1$
\newauthor{Yun Liu$^1$, Wenqaing Deng$^1$, Zuhui Fan$^1$\thanks{E-mail: zuhuifan@ynu.edu.cn}}\\
$^1$ South-Western Institute for Astronomy Research, Yunnan University, Kunming, 650500, P. R. China; \\
$^2$Yunnan Observatories, Chinese Academy of Sciences, Kunming, 650216, P. R. China;\\
$^3$The Shanghai Key Lab for Astrophysics, Shanghai Normal University, Shanghai, 200234, P. R. China;
}
\date{Accepted XXX. Received YYY; in original form \today}
\pubyear{2020}

\begin{document}
\label{firstpage}
\pagerange{\pageref{firstpage}--\pageref{lastpage}}
\maketitle
\begin{abstract}
    Convolutional Neutral Networks have been successfully applied in searching for strong lensing systems, leading to discoveries of new candidates from large surveys.
    On the other hand, systematic investigations about their robustness are still lacking. In this paper, we first construct a neutral network, and apply it to $r$-band images of Luminous Red Galaxies (LRGs) of the Kilo Degree Survey (KiDS) Data Release 3 to search for strong lensing systems. We build two sets of training samples, one fully from simulations, and the other one using the LRG stamps from KiDS observations as the foreground lens images. With the former training sample, we find 48 high probability candidates after human-inspection, and among them, 27 are newly identified. Using the latter training set, about 67\% of the aforementioned 48 candidates are also found, and there are 11 more new strong lensing candidates identified. We then carry out tests on the robustness of the network performance with respect to the variation of PSF. With the testing samples constructed using PSF in the range of 0.4 to 2 times of the median PSF of the training sample, we find that our network performs rather stable, and the degradation is small. 
    We also investigate how the volume of the training set can affect our network performance by varying it from 0.1 millions to 0.8 millions. The output results are rather stable showing that within the considered range,  our network  performance is not very sensitive to the volume size.
    
\end{abstract}


\section{Introduction}
The images of distant galaxies are distorted by the tidal effect of the gravitational potential
generated by intervening matter: an effect commonly referred to as gravitational lensing 
\citep[see, e.g.][for a detailed introduction]{Schneider2006}.
Gravitational lensing is one of the {powerful probes}
in studying the dark Universe as well as the galaxy formation and evolution \citep[e.g.][]{Bonvin2018,Hildebrandt2017}. In particular, the strong gravitational lensing can produce {arc-shaped} distortions or multiple images of one or more background sources \citep[e.g.][]{Treu2010}. Such systems can provide important cosmological constraints, such as on the equation of state of dark energy \citep{Collett2014}, and the Hubble constant \citep[e.g.][]{Suyu2017}. They can also be used to measure the mass distribution of the central region of lens galaxies \citep[e.g.][]{Luo2018,Prat2018,Prat2018-mn}, and to probe the dark substructures \citep[e.g.][]{Mao1998, Xu2009}. Moreover, a strong lens can act as a "cosmic telescope", to magnify the faint background sources, and thus provide a unique tool to study the extremely high redshift {faint objects}, which {cannot} be visible or resolved otherwise \citep[e.g.][]{2007ApJ...666...45L, 2012ApJ...745..155H,Kelly2018}.

Since the discovery of Q0957+561 by \cite{Walsh1979}, over a few hundreds of galactic-scale strong lensing systems have been confirmed \citep{Treu2010, Shu2016}. Different searching schemes have been applied, including visual inspection of deep optical images \citep{Fevre1988,Sygnet2010,Pawase2014,Diehl2017,Nord2020}, targeted observations of potentially lensed quasars and the sub-millimeter galaxies in the bright end of the luminosity function \citep{Negrello2010}, and follow-up of systems for which optical spectroscopy revealed anomalous emission lines \citep[e.g.][]{bolton2004,Richard2011,Shu2017}. The method of using time-domain information has also been proposed as an efficient approach \citep{pindor2005,Kochanek2006}. 

So far the largest sample of confirmed strong lensing systems is the Sloan Lens Advanced Camera for Surveys \citep[SLACS, ][]{bolton2006,bolton2008}, which contains more than 100 systems identified from SDSS (Sloan Digital Sky Survey) spectral features and then confirmed by high resolution imaging. The on-going third generation gravitational lensing imaging surveys, including KiDS\footnote{https://kids.strw.leidenuniv.nl}, DES\footnote{https://www.darkenergysurvey.org} (Dark Energy Survey) and HSC\footnote{https://hsc.mtk.nao.ac.jp/ssp/survey} (Hyper Suprime-Cam), will be able to yield thousands of new lenses \citep{deJong2013,Jacobs2019b,Sonnenfeld2018}. Furthermore, the future wide-field imaging surveys such as Euclid\footnote{https://www.euclid-ec.org}, LSST\footnote{https://www.lsst.org} (Large Synoptic Survey Telescope), WFIRST\footnote{https://wfirst.gsfc.nasa.gov} (Wide Field Infrared Survey Telescope) and CSST (Chinese Space Station Telescope), can increase the number of {detected} strong lensing systems by orders of magnitude {\citep[see, e.g.][]{Euclid-intro,Oguri2010,Marshall2010,WFRIST-intro,CSST-intro,Gong2019}}. To search for strong lensing systems from millions to over a billion of galaxy images is a challenging task. It is important to develop robust and efficient algorithms. In \citet{Metcalf2019}, a data challenge for detecting strong lensing systems was presented. Several teams using algorithms ranging from human-inspection to Convolutional Neural Networks (CNN) submitted the results. {Although the analyses show} that the algorithms based on human or computer have different advantages, the computer-based algorithms generally received higher scores showing a promising potential in strong lensing detection from big data sets.

The Artificial Intelligence (AI) technology opens a new window in astronomical studies, and has been widely applied to galaxy classification \citep{Lukic2018,Pérez-Carrasco2019}, photometric redshift measurement \citep{Cavuoti2017,Pasquet-Itam2018}, supernova classification\citep{Cabrera-Vives2017,Pasquet2019a}, image de-blending \citep{Boucaud2019} and lens modelling \citep{Pearson2019}. One of the most successful applications is in the area of strong lensing search \citep{Petrillo2017,Ostrovski2017,Schaefer2018,Ma2018,Davies2019,Jacobs2019a,Petrillo2019,Li2020,2020arXiv200413048C}. To meet the demands of future large surveys, however, there are still issues remaining to be explored. For examples, how does the volume size of training samples impact the performance of a network? How does a specific network perform under different observational conditions, e.g. with different Point Spread Function (PSF)? These are non-trivial problems and deserve careful studies.

In this paper, we carry out CNN studies in searching for strong lensing systems. We describe the architecture of our Network (hereafter, we use the capitalised Network for our specific network) in Section\,\ref{sec:Network}, and present the construction of training samples in Section\,\ref{sec:training_samples}. In Section\,\ref{sec:lensing_candidates}, we apply our Network to KiDS data, and show the 27 newly identified candidates. In Section\,\ref{sec:test_the_network}, we investigate the impacts of the PSF variation and the size of training samples on the performance of our Network. Summary and discussions are shown in Section\,\ref{sec:summary}. In this paper, we adopt the standard $\Lambda$CDM cosmology with parameters: $\Omega_\Lambda = 0.7$, $\Omega_m = 0.3$ and $\Omega_k = 0$.

\begin{figure}
    \centering
    \includegraphics[scale=0.5]{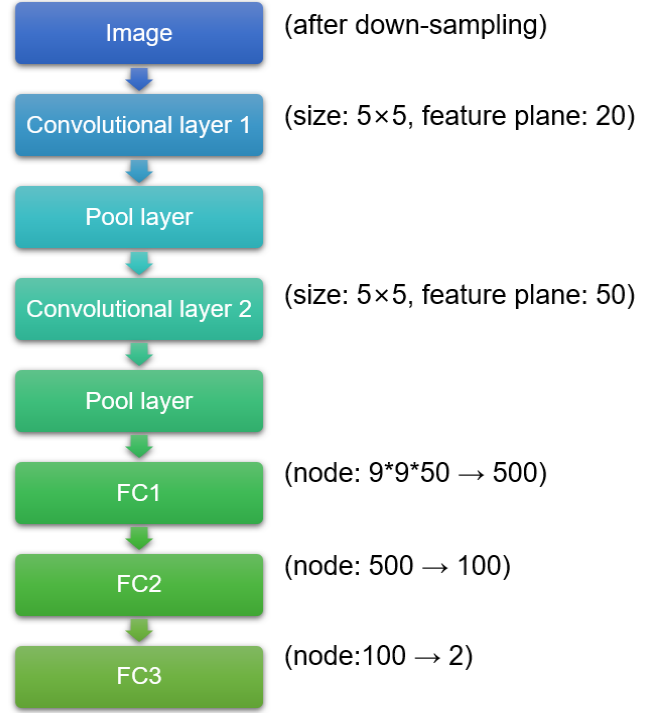}
    \caption{The architecture of {our Network}. `FC' { stands for} full connection layer.}
    \label{fig:network_structure}
\end{figure}

\section{Neural Network}
\label{sec:Network}

Deep neutral networks can have very different architectures, depending on the problems to be solved in different domains or use-cases. CNN \citep{Krizhevsky2012} is inspired by the human visual cortex and is the neural network of choice specially for image and video recognition. 

A typical CNN consists of a series of convolutional and pooling layers followed by a full connection layer and a normalising (e.g., softmax function) layer. The convolutional layers can maintain the spatial continuity of an image and extract its local features. The pooling layers may use max pooling or mean pooling to reduce the dimensions of the middle hidden layers, and the amount of operations of the next layers, and to provide translation invariance.

More formally, a typical neural network can be expressed as: $x_{l+1,i} = f\left(\sum_{i'}{w_{l,i,i'}x_{l,i'}+b_{l,i}}\right)$, where $l$ is the neural layer index, $i$ is the output neuron index and $i'$ is the input neuron index, $x_{l,i'}$ is the input of layer $l$ and {$x_{l+1,i}$} is the output of layer $l$ (also the input of layer $l+1$), $f$ is an activation function, $w_{l,i',i}$ is the weights of layer $l$ and $b_{l,i}$ is the bias of layer $l$. For CNNs, the form of convolution becomes $x_{l+1,c,i} = f\left(\sum_{c',i'}{w_{l,c,c',i'}x_{l,c',i'}+b_{l,c}}\right)$, where $c$ is the output channel index and $c'$ is the input channel index. In contrast to the normal neural network, each output neuron in convolutional layers is connected to the input neurons locally instead of being connected to all input neurons.

There are some significant benefits of using CNNs in searching for strong lensing systems. First, CNNs' structure is suitable to capture the features of gravitational lensing systems, such as the stationarity of statistics and the locality of pixel dependencies. Secondly, the capacity of CNNs can be easily controlled by varying their depth and breadth in order to search for variety of lensing systems. Thirdly, CNNs have been proven to be powerful in image classifications. Finally, there are efficient CNN implementations both in CPU and GPU.

Because of these benefits, in this study, we construct our Network for strong lensing detection mainly based on the structures of CNNs. Specifically, as shown in Figure\,\ref{fig:network_structure}, our input data is 16 bits 49$\times$49 one channel image. The 1st layer is a 5$\times$5 convolutional layer with 1 input channel and 20 output channels. The 2nd layer is a 2$\times$2 max pooling layer with 2$\times$2 strides. The 3rd layer is another 5$\times$5 convolutional layer with 20 input and 50 output channels. The 4th layer is again a 2$\times$2 max pooling layer with 2$\times$2 strides. The 5th layer is a full connection layer consisting of 4050 inputs and 500 outputs. The 6th and 7th layers are again full connection layers, with 500 inputs and 100 outputs, and 100 inputs and 2 outputs, respectively. We use the ReLU activation function for every neuron in layer 1,3,5 and 6, and no activation function is used for the other layers. 

Compared to other architectures, our Network has the following features:

1. The structure is relatively simple, especially compared with some very deep network such as ResNet \citep{He2016}, which normally needs residual network structures to make the training feasible. 
 
2. Due to the concise structure, the number of parameters is much fewer in our Network than in some complicated CNNs. We have $2,101,372$ parameters, which is only about 3\% of 60M parameters in ResNet. Fewer parameters will help to prevent over-fitting and improve interpretability.

3. Our Network can be trained by relatively small data sets, and thus can be run in a typical desktop. This makes it easy to use.

We design this relatively shallow network structure because strong lensing systems typically have much fewer features than the images, e.g., the ones in ImageNet, handled by the complicated ResNet \citep{JiaDeng2009}. To quantitatively compare the complexity of our training samples and those in ImageNet, we compute the Shannon entropy \citep[entropy hereafter,][]{Hammer2000,Shan2008} of our simulated training samples (Section\,\ref{sec:training_samples}) and an officially published subset of images in ImageNet.\footnote{http://image-net.org/small/download.php} This subset contains 49,999 $64\times64$ RGB images, with the format of uint8 (8-bit unsigned integer). To meet its resolution and data format, we lower the resolution of our simulated samples and convert our uint16 (16-bit unsigned integer) data to uint8. Additionally, we randomly select 50,000 images with half positive and half negative from our training sample (see Section\,\ref{sec:training_samples} to find further details of training sample). Before calculating the entropy, the RGB images from ImageNet are converted to grey-scale images through the following formula:
\begin{equation}
    I_{grey} = 0.2989 * R + 0.5870 * G + 0.1140 * B     
\end{equation}
where $R,G,B$ denote the pixel values of the respective bands.

The entropy for each image is calculated as follows \citep{Hammer2000}:
\begin{equation}
    E(\gamma)=-\Sigma p_i {\rm log_2} p_i
\end{equation}
Here $\gamma$ represents an image with different discrete grey-scale pixel values. For unit8 images here, there are $2^8=256$ pixel values (denoted as $g_i$, $i=1, ..., 256$). The summation in the right side goes through all the $256$ values, and $p_i$ is the number of pixels with values $g_i$ normalised to the total number of pixels. It is seen that the more variety of $g_i$ values is, the more uniform of $p_i$ is, leading to a larger entropy. In other words, a larger entropy indicates a higher complexity of an image. 

We calculate the entropy values for the 49,999 ImageNet images and 50,000 our training images, respectively. The average entropy value for ImageNet is $7.11\pm 0.66$, where the error is the standard deviation of the entropy values of the images in this set. For our simulated training samples, the value is $4.24 \pm 1.49$ for positive samples and $4.07 \pm 1.43$ for negative ones. It is seen that the entropy of our training sample is $\sim58\%$ of the value from the images in ImageNet. This indicates that our classification problem in strong lensing search is simpler than that for ImageNet. Correspondingly, to train a network aiming at strong lensing detection, the training sample is typically smaller than that of ImageNet. For example, the Strong Gravitational Lens Finding Challenge \citep[Lens Finding Challenge,][]{Metcalf2019} has only $100,000 \times 4$ images for training, compared with the order of 50 million cleanly labelled full resolution images in ImageNet. In general, a deep architecture model suffers from over-fitting problem when there is a small number of training data \citep{Pasupa2016}. The bias-variance trade-off of supervised learning also suggests that a simple architecture model is preferred over a complex model if data are "simple" \citep{Luxburg2011}. This is how Occam's razor principle works in the area of machine learning. This guideline has been proved by some experiments \citep{Schindler2016}, which showed that shallow models are more appropriate for relatively small and simple data sets.

It is noted that our Network is AlexNet-based \citep{Krizhevsky2012}, with two convolutional layers followed by three full connection layers. Similar architectures have been adopted by other groups in their strong lensing search codes, e.g., AstrOmatic, GAMOCLASS and NeuralNet2 \citep{Davies2019}. These groups participated Lens Finding Challenge. From Table 3 of \cite{Metcalf2019}, we can see that their performance is comparable with ResNet-based methods, such as CMU DeepLens \citep{Ma2018} and Kapteyn Resnet \citep{Petrillo2019b}. Moreover, these networks are ranked close to LASTRO EPFL \citep{Schaefer2018}, which is also an AlexNet-based network but with 5 convolutional layers. We therefore conclude that the architecture of our Network is suitable for strong lensing searches. Meanwhile, we expect that our testing results shown in Section\,\ref{sec:test_the_network} should be instructive to other studies using similar network structures. 

We train the Network by using Adam algorithm \citep{Kingma2015}, which is an adaptive learning rate optimisation algorithm. It has been designed specifically for training neural networks and can be considered at as a combination of RMSprop and Stochastic Gradient Descent with momentum. The weight of every neuron $w$ will be updated in the training as follows. Denote $g_t$ as the gradient of $w$ at the iteration $t$, $m_t$ as the estimate of mean of $g_t$ and $v_t$ as the estimate of uncentred variance of $g_t$, then we can compute them by the decaying moving average method as:
\begin{gather}
m_t=\xi_1m_{t-1}+(1-\xi_1)g_t \\
v_t=\xi_2v_{t-1}+(1-\xi_2)g_t^2
\end{gather}
Because $m_t$ and $v_t$ are initialised as zeros, they are biased towards zeros. These biases can be calculated with the following steps:
\begin{gather}
\hat{m_t}=\frac{m_t}{1-\xi_1^t} \\
\hat{v_t}=\frac{v_t}{1-\xi_2^t}
\end{gather}
Then we can update $w$ as followed:
\begin{equation}
w_{t+1}=w_t-\frac{\eta}{\sqrt{\hat{v_t}}+\epsilon}\hat{m_t}
\end{equation}
where $\eta$ is learning rate, $\xi_1$ is the exponential decay rate for $m$, $\xi_2$ is the exponential decay rate for $v$ and $\epsilon$ is a very small number to prevent any division by zero in the implementation. Normally, $\xi_1$, $\xi_2$ and $\epsilon$ have really good default values of 0.9, 0.999, and $10^{-8}$ respectively \citep{Kingma2015}. The learning rate $\eta$ is the proportion that weights are updated. Larger values result in faster initial learning rate. Smaller values slow learning rate down during training. With a large $\eta$, it is highly possible that the neural network quickly converge to a sub-optimal solution. While a small value of $\eta$ can stuck the training process. For our Network, we tune the value of $\eta$ experimentally based on the Network outputs (see Section\,\ref{sec:size tst}).

The way to use the training data is mainly controlled by the number of epochs and the batch size. One epoch means that the entire data set is passed forward and backward through the neural network once. {Because typically the data set is too big to be fed to the computer at once}, normally the entire data is divided into several smaller batches. The batch size and the total number of epochs can influence the final training result significantly. Therefore we also tune these two parameters during our training {(see Section\,\ref{sec:size tst})}. 

Our Network outputs a probability for an image being a strong lensing system. The negative log-likelihood is employed to measure the loss. We use the accuracy and the completeness to evaluate the performance of the Network. We denote TP for true positive, FP for false positive, TN for true negative, and FN for false negative. The accuracy is defined as (TP+TN)/(TP+TN+FP+FN), which measures how well our Network can correctly identify lensing/non-lensing systems. The completeness is defined to be TP/(TP+FN), reflecting the capability of our Network to find lensing systems without mis-classifying them as non-lensing systems. We also use the false positive ratio FPR=FP/(FP+TN) to measure the performance of our Network. The smaller the FPR is, the better the performance is.

\section{Training Samples}
\label{sec:training_samples}
For strong lensing detection using CNNs, besides the learning capability of a network, the training sample employed to train the network can also affect the accuracy and completeness significantly. The training set is necessary to sample the data to be investigated as fairly as possible. Because the currently confirmed strong lensing systems are insufficient to represent fully the new data to be studies, thus simulations are often used to generate training data \citep[e.g.][]{Ma2018,Sonnenfeld2018}. Here we adopt the simulation approach to prepare our training sample as well. 

In order to train our Network, we construct a large sample of galaxy images, which consists of both lensing images (positive sample) and non-lensing images (negative sample). In this study, we focus on KiDS-like surveys, and thus the simulation settings are in accord with that of KiDS observations\footnote{http://kids.strw.leidenuniv.nl/DR3/index.php} \citep{DeJong2017}. We describe how we prepare the images in the training sample in this section.

\subsection{Lensing Basics}

The theory of gravitational lensing can be found from general reviews, e.g. \cite{BARTELMANN2001291} and \cite{Treu2010}. We outline the basic formalism here. In this study, we only consider cases with a single lens and adopt the thin lens approximation. The deflections of light rays occur as they pass through the lens plane between the source and observer. The relationship between the source position  
$\beta=\sqrt{\beta_x^2 + \beta_y^2}$ and the observed image position $\theta=\sqrt{x^2 +y^2}$ is given by the lens equation
\be
\beta = \theta -\frac{D_{ds}}{D_s}\hat{\alpha},
\ee
where $\hat{\alpha}$ is the deflection angle determined by the mass distribution of the lens, and $D_{ds}$ and $D_s$ are the angular diameter distances from lens to source, and from observer to source, respectively. 

\subsection{Positive Samples}
\label{kids, training set, pos samples}
To be consistent with the searches around LRGs of KiDS survey, we adopt the elliptical galaxies as our lenses. We model their mass distribution by the Singular Isothermal Ellipsoid \citep[SIE,][]{Kormann1994}, which has been shown to be in a good agreement with observations \citep[see e.g.][]{Bolton12, Sonnenfeld13}. The surface mass density scaled by the lensing kernel, referred to as the lensing convergence, for an SIE halo can be written as 
\be
\kappa(x,y)=\frac{\theta_E}{2}
\dfrac{1}{\sqrt{qx^2 +y^2/q}},
\label{eq:lensing potential}
\ee
where 
\be
\theta_E=4\pi (\frac{\sigma}{c})^2 \frac{D_{ds}}{D_s}
\elabel{einsteinradius}
\ee
is the Einstein radius, $\sigma$ is the velocity dispersion of the SIE halo, and $q$ is the axial ratio between the minor and major axes. In our simulation, $q$ is chosen to be uniformly distributed between 0.7 to 1 \citep{Clampitt2016}. The axial ratio of the luminous lens galaxy is assumed to be the same as that of its host halo.

\begin{figure}
    \centering
    \includegraphics[scale=0.5]{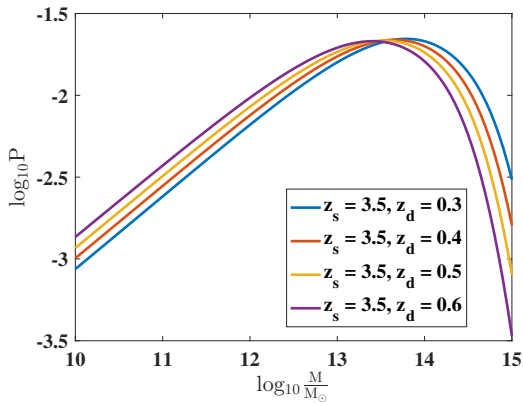}
    \caption{Normalised lensing probability of four redshift combinations, given by Equation\,\ref{eq:prob}.}
    \label{fig:lens_prob}
\end{figure}

\begin{figure}
    \centering
    \includegraphics[scale=0.5]{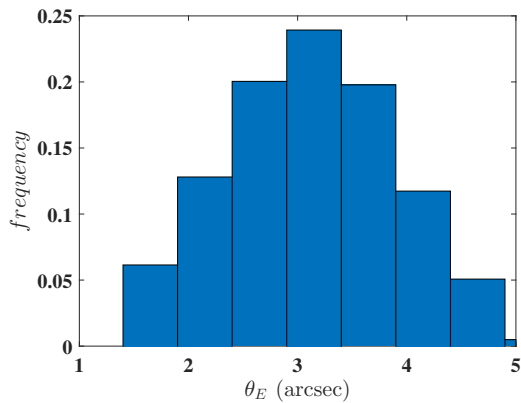}
    \caption{Histogram of $\theta_E$ of the 100,000 simulated lenses. In the simulations, we consider $\theta_E$ from 1.4 to 5 arcsec.
    }
    \label{fig:theta_e}
\end{figure}

We simulate lensing systems according to the following probability:
\begin{equation}
    {\rm P}(M|z_s, z_d) = p(M|z_d){\theta_E}^2(M|z_s, z_d)
    \label{eq:prob}
\end{equation}
where $M$ is the mass of a lens halo; $z_s$ and $z_d$ are the redshifts of the source and the lens, respectively. The redshift combinations of the lens and source we used are listed in Table\,\ref{redshift combination}, and $p$ is the halo mass function with given $M$ and $z_d$. We use the halo-mass function from \cite{Reed2007}. We show examples of the lensing probability for 4 redshift combinations in Figure\,\ref{fig:lens_prob}. Figure\,\ref{fig:theta_e} shows the distribution of the Einstein radius of our simulated lenses.

In addition to the main halo, we also include subhaloes in our simulations, while the external shear is not taken into account. The substructures can induce changes of the flux ratio between the multiple images and generate small image distortions. We use the Singular Isotherm Sphere (SIS) to model the mass distribution of the subhaloes. The 
mass function of subhaloes is taken from \cite{Xu2009},
\begin{equation}
	\frac{ d\ \eta_{sub} }{d \ ln\ m_{sub}}= 0.01(\frac{m_{sub}}{3 \times 10^8h^{-1}\ M_\odot})^{-1}(h^{-1}kpc)^{-2}.
	\label{eq:sub-massfunc}
\end{equation}
where $m_{sub}$ is taken in the range from $1\%$ to $5\%$ of $M_{200}$ of the main halo. If the lower limit is smaller than $10^9 \hbox {M}_{\odot}$, we take it to be $10^9 \hbox {M}_{\odot}$. Smaller subhaloes are not included because we do not expect to observe their effects on the lensing images for ground-based KiDS-like observations. The upper mass limit of the subhaloes we choose is relatively small. The reason is that, in this study, we mainly concern smoothed LRGs as foreground lenses. Large subhaloes may host satellite galaxies and thus induce complexities the LRG images. We will see later that in searching for strong lensing candidates from KiDS DR3, we perform a pre-selection and only search for lensing systems around foreground LRGs. We visually inspect the LRG images, and find that the probability containing a large luminous clump within $5''$ around LRGs is low. We therefore do not include larger subhaloes in our simulations. It is noted that in our analyses, all the considered subhaloes only contribute to the mass distribution, but not to the light distribution of the foreground images. We populate subhaloes randomly within their main halo without taking into account the profile of their spatial distribution. Because in this paper we focus on finding strong lensing candidates rather than lensing modelling, the spatial distribution profile of the subhaloes do not affect our results significantly. 

We note that in our simulations, we do not take into account line-of-sight structures. Such structures can contribute to the lensing effects, as well as contaminate the foreground optical images \citep[e.g.][]{xu2012,Despali2018}. For strong lensing detections in future deep surveys, the line-of-sight effects should be carefully considered. Also, a more sophisticated modelling about the afore discussed substructures will be necessary.

We calculated $M_{200}$ of a main halo by \citep{1Mo1998}
\begin{equation}
 M_{200} = \dfrac{\sigma^3}{10GH(z_d)},
\end{equation}
where $G$ is the gravitational constant and $H(z_d)$ is the Hubble parameter at the lens redshift $z_d$. The radius $R_{200}$ can be calculated correspondingly. For each host halo, we consider 10 subhalo mass bins to calculate the numbers of subhaloes. The subhaloes are randomly distributed within the range of $R_{200}$ of the host halo. 

For the light distributions of lens and source galaxies, we employ the Sersic profile \citep{Sersic1963} to model their brightness:
\be
I_s(\theta) = I_e\exp(-b_n((\frac{\theta}{\theta_e})^\frac{1}{n}-1)),
\ee
where $I_e$ is the intensity at $\theta_e$, $n$ is the Sersic index, $b_n=1.9992n - 0.3271$ and $\theta_e$ is the effective radius. We link the luminosity $L$ of a galaxy and $\sigma$ of its corresponding SIE halo by the Faber-Jackson relation \citep{FJ1976}, 
\be
L \propto {\sigma}^4.
\ee
In our simulations, we adopt the proportional coefficient in this relation given by \cite{schneider2014extragalactic}. Using the spectral energy distribution (SED) and the redshift of a lens galaxy, its apparent magnitude in $r-$band can be calculated. For the effective radius of lens galaxies, we use the scaling relation between the Einstein radius and the effective radius from \cite{LiRui2018}. The Sersic index is uniformly distributed between 2 and 5 in our simulations, which is within the typical range for elliptical galaxies \citep{Huertas-Company2013}. Moreover, we assume that the orientation of a lens galaxy coincides with that of its {host halo} with the same position angle ($P.A$.), which is evenly distributed between 0 and $180^o$. 

\begin{figure}
    \centering
    \includegraphics[scale=0.5]{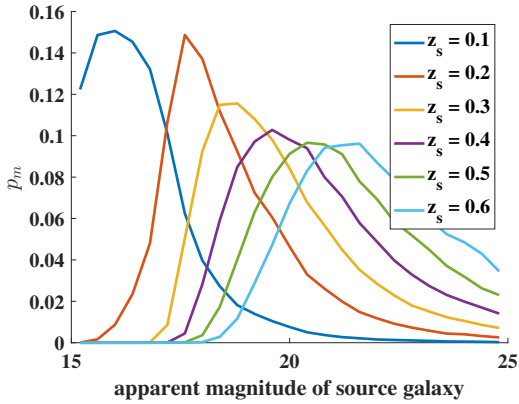}
    \caption{The distributions of apparent magnitude of source galaxies at six redshift bins. Here $p_m$ is the frequency of the number of galaxies in each magnitude.}
    \label{fig:app_mag_dis_kids}
\end{figure}

For the source galaxies, we follow the distribution of apparent magnitude at certain redshifts given by \cite{Benitez2000}. Examples are shown in Figure\,\ref{fig:app_mag_dis_kids}. We assign the effective radius randomly between 0.2 and 0.7 arcsec and the Sersic index randomly between 2 and 5. The Sersic index range for source galaxies here is somewhat biased to early-type galaxies, and thus may not be a full representation of the observed source galaxy sample. In this study, we consider only single $r$-band images without incorporating colour information. Thus the lack of late-type galaxies in our simulated source galaxy sample may not affect the strong lensing search significantly. As we discuss later in Section\,\ref{sec:summary}, in our future studies, we intend to include galaxy colours into analyses, and we will then consider a broader range of $n$ that is correlated with galaxy colours in our simulations to better resemble source galaxies.

We note that in our simulations, the lens halo mass function and the galaxy magnitude distribution are taken from theoretical modelling and observational modelling, respectively, and they are not flatly distributed. On the other hand, for parameters related to the galaxy light distribution, such as $q$, $n$ and $\theta_e$, we adopt flat distributions within reasonable parameter ranges. The flat distributions of these parameters are also adopted in other studies, e.g., in \cite{Petrillo2017}. In Section\,\ref{subsec: 4.3.3}, we test the Network performance using a different distribution for the Sersic index $n$. It shows that the performance is not sensitive to the specific distributions as long as the parameter ranges are similar. Thus in this paper, our default setting is flat distributions for these parameters. 
There are observational analyses showing the distributions of these parameters \citep[see e.g.][]{Griffith2012,Huertas-Company2013,Roy2018}. Those results however focus on a specific population of galaxies, and the results still have large uncertainties. At the moment, it is not easy to make simulations with the parameters distributions accurately resembling the real data to be analysed. We do not expect a significant impact to current stage of study. However, in future work targeting at high precision analyses, image simulations with more careful settings for the galaxy parameters are needed.

In order to simulate images with significant strong lensing features, source galaxies are placed near caustics on the source plane. Specifically, the offset between the centre of a source galaxy and the caustics is smaller than the effective radius of the source galaxy. The relevant parameters that we use are summarised in Table\,\ref{parameters, kids, pos}. 

\begin{figure*}
\centering
    \begin{minipage}{0.2\linewidth}
    \centerline{\includegraphics[width=1\textwidth]{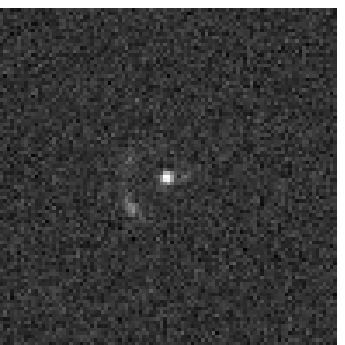}}
    \centerline{S/N=20.41}
    \hspace{0.15in}
    \end{minipage}
\qquad
    \begin{minipage}{0.2\linewidth}
    \centerline{\includegraphics[width=1\textwidth]{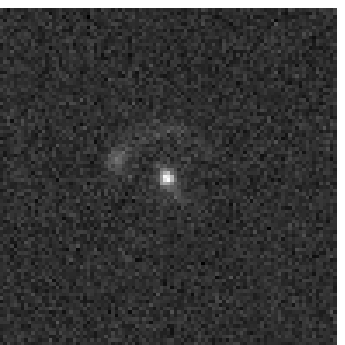}}
    \centerline{S/N=29.40}
    \hspace{0.15in}
    \end{minipage}
\qquad
    \begin{minipage}{0.2\linewidth}
    \centerline{\includegraphics[width=1\textwidth]{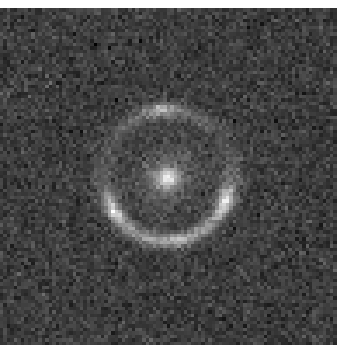}}
    \centerline{S/N=49.40}
    \hspace{0.15in}
    \end{minipage}
\qquad
    \begin{minipage}{0.2\linewidth}
    \centerline{\includegraphics[width=1\textwidth]{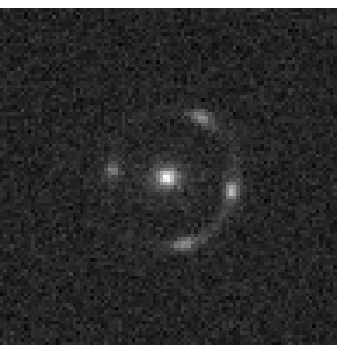}}
    \centerline{S/N=48.16}
    \hspace{0.15in}
    \end{minipage}
\caption{Examples of positive images under the KiDS conditions. The size of the image stamp is 97 $\times$ 97 pixels, which gives us about 20 $\times$ 20 arcsec$^2$. 
{ The corresponding S/N of each images is also given at the bottom.}}
\label{fig:eg_pos_samples} 
\end{figure*}

\begin{table}
    \centering
\begin{tabular}{cccccc}
\hline
\multicolumn{2}{c}{$q$} & \multicolumn{2}{c}{$\theta_E\space({\rm arcsec})$} & \multicolumn{2}{c}{$\theta_e\space({\rm arcsec})$} \\ \hline
lower      & upper      & lower              & upper             & lower              & upper             \\
0.7        & 1.0        & 1.4                & 5                 & 0.2                & 0.7               \\ \hline
\multicolumn{2}{c}{$n$} & \multicolumn{2}{c}{$mag_s$}            & \multicolumn{2}{c}{$\phi\space({\rm degree})$}     \\ \hline
lower      & upper      & lower              & upper             & lower              & upper             \\
2          & 5          & 15                 & 25                & 0                  & 180               \\ \hline
\end{tabular}
\caption{The parameters that are used in the fully simulated positive sample: $q$ is the axial ratio of lenses and sources; $\theta_E$ is the Einstein radius of the lensing systems; $\theta_e$ is the effective radius of source galaxies; $n$ is the Sersic index of lens and source galaxies; $mag_s$ is the apparent magnitude of source galaxies; $\phi$ is the position angle of lens galaxies and their main haloes. 
}
\label{parameters, kids, pos}
\end{table}

\subsection{Negative Samples}
\label{parameters, kids, neg}
A representative negative sample also plays an important role in training our Network. As our purpose is to detect strong lensing candidates efficiently, we need to include different cases in the negative sample to reduce the FPR as much as possible. To enhance the ability of our Network, we prepare different types of galaxies in the negative sample. The parameter ranges for the galaxies are set in accord with \cite{Griffith2012,Miller2013, Shibuya2015}. Flat distributions are used again for the parameters of galaxy images.

In addition to single regular galaxies in the negative sample, we also generate cases of two elliptical galaxies, with one in the centre of the image stamp and the other one randomly placed in the stamp. This can reduce the FPR of galaxy pairs effectively. We also create galaxy images with extreme axial ratios ($q \in\,[0.4,0.5]$). We include them in the negative sample to further reduce the FPR because to a certain extent, the elongated galaxies may show similar shape as the lensed arc-images. For all galaxies in the negative sample, we use the Sersic model for their light distributions. For elongated galaxies, the adopted range of $\theta_e$ is broader. Because in our testing results, the low axial ratio and large $\theta_e$ galaxies are two kinds of galaxies bewildering our Network.

We also include disk galaxies containing a Sersic bulge component and a Sersic disk component in the negative sample. The luminosity ratio of the two is uniformly distributed between 0.2 to 3. The effective radii of the two components are both taken to be in the range from 0.2 arcsec to 2 arcsec. The axial ratios of the two components are different: the bulge is more spherical with $q\in [0.95,1.0]$, while the disk has a large range with $q\in [0.4,1.0]$. The disk orientation is randomly distributed. The parameters ($q$, $\theta_e$ and $n$) of all kinds of galaxies that we include in the negative sample are listed in Table\,\ref{parameters, kids, neg12}.

There are equal numbers of positive and negative images in the training sample. The negative ones consist of 80\% elliptical galaxies (half of them are paired galaxies), 10\% highly elongated galaxies and 10\% disk+bulge galaxies.

\begin{table*}
    \centering
\begin{tabular}{ccccccc}
\hline
                                       & \multicolumn{2}{c}{$q$} & \multicolumn{2}{c}{$\theta_e(arcsec)$} & \multicolumn{2}{c}{$n$} \\ \hline
\multirow{2}{*}{elliptical galaxies}   & lower      & upper      & lower              & upper             & lower      & upper      \\ \cline{2-7} 
                                       & 0.7        & 1.0        & 0.2                & 0.7               & 2          & 5          \\ \hline
\multirow{2}{*}{elongated galaxies}    & lower      & upper      & lower              & upper             & lower      & upper      \\ \cline{2-7} 
                                       & 0.4        & 0.5        & 0.2                & 1.7               & 2          & 5          \\ \hline
\multirow{2}{*}{disk galaxies (disk)}  & lower      & upper      & lower              & upper             & lower      & upper      \\ \cline{2-7} 
                                       & 0.4        & 1.0        & 0.2                & 2                 & 0.9        & 1.1        \\ \cline{2-7} 
\multirow{2}{*}{disk galaxies (bulge)} & lower      & upper      & lower              & upper             & lower      & upper      \\ \cline{2-7} 
                                       & 0.95       & 1.0        & 0.2                & 2                 & 1.5        & 5          \\ \hline
\end{tabular}
\caption{The parameters of fully simulated negative sample.}
\label{parameters, kids, neg12}
\end{table*}

\subsection{PSF and Noise}
To account for the observational seeing conditions, we smear the images with a PSF. We choose a Gaussian function as the PSF kernel
\begin{equation}
P(\theta) = \frac{1}{{\sqrt{2\pi}} {\sigma_p}}\exp(\frac{-{\theta}^2}{2\sigma_p^2}),
\end{equation}
where $\sigma_p$ is related to the FWHM seeing by $\sigma_p= {\rm FWHM}/2.355$. 
{For KiDS, the FWHM is about $0.65''$ \citep{DeJong2017}. We thus take FWHM to be in the range of $0.55''$ to $0.75''$ in our simulations. In Section\,\ref{sec:psf tst}, we test the Network performance on data sets with a broader range of PSFs.}

A Poisson noise is added to each pixel, with the mean taken to be the intensity of the pixel. Finally, the CCD read-out noise and the sky background are added using a Gaussian model. To decide the Signal-to-Noise ratio (S/N), we fit empirical relationship between the apparent magnitude and the S/N from the catalogue of KiDS DR3 \citep{DeJong2017}, and obtain an approximate expression 
\begin{equation}
{\rm S/N}({mag})=
\begin{cases}
a\exp{(b*mag)}& {mag<23}\\
15& {else}
\end{cases},
\label{eq:mag2sn}
\end{equation}
where $a = {\rm 1.796E+09}$, $b = -0.8079$ and $mag$ is the apparent magnitude of the foreground galaxies.
To account for scatters, the S/N adopted in our simulations has a range from 0.9 to 1.1 times the S/N value calculated by Equation \ref{eq:mag2sn}. The noise is added in the same way for both positive and negative images. A few examples of positive images are shown in Figure\,\ref{fig:eg_pos_samples}.

The pixel size of the simulated images is $0.21''$, in accord with that of KiDS. The size of the image stamps is 97 $\times$ 97 pixels, which is about 20 $\times$ 20 $\rm{arcsec}^2$. In our fiducial setting, the volume size of the training sample is 100,000 images, half positive and half negative.

\section{Apply to The KiDS}
\label{sec:lensing_candidates}
KiDS DR3 \citep{DeJong2017} covers 255 square degrees and contains 292 survey tiles, including single-band and multi-band catalogues ($u$,$g$,$r$ and $i$). In this paper, we concentrate on searching for strong lensing candidates from KiDS $r$-band images, which have the best quality among the four bands with the median PSF of 0.65 arcsec. With the training sample described in Section\,\ref{sec:training_samples}, we apply our trained Network to KiDS DR3 $r$-band image data. Similar to \cite{Petrillo2017} and \cite{Petrillo2019}, we search for strong lensing candidates around LRGs only. For that, we construct the LRG sample with the following criteria.
We first select the extend sources as follows:
\begin{flushleft}
    \romannumeral1. exclude all stars. We select the sources with \mbox{{\tt CLASS\_STAR}} between 0.01 to 0.03;\\ 
    \romannumeral2. the sources with \mbox{{\tt FLUX\_RADIUS\_R}} 5 to 30 arcsecs;\\
    \romannumeral3. the sources have detection in all four bands: u,g,r,i.\\
\end{flushleft}
After this selection, we have 818,384 extend sources. In the second step, we select LRGs using the same colour criteria as that in \cite{Petrillo2017}:
\begin{center}
    $r<20$\\
    $|c_{perp}|<0.2$\\
    $r<c_{par}/0.3$,
\end{center}
where 
\begin{center}
    $c_{par}=0.7(g-r)+1.2[(r-i)-0.18]$\\
    $c_{perp}=(r-i)-(g-r)/4.0-0.18$.
\end{center}
In the end, we have 28,815 LRGs as our KiDS test sample. Our sample is larger than that used in \cite{Petrillo2017}, which contains 21,789 LRGs, mainly because they further exclude sources containing masked regions within stamps.

We apply the same procedure and further limit to the sources with \mbox{{\tt FLAG\_R}} $ = 0$, i.e., no masks. The number of LRGs is then reduced to 23,067. The other minor difference between our selection and that in \cite{Petrillo2017} is the criteria applied to find extended sources. In \cite{Petrillo2017}, two conditions are applied: 1) \mbox{{\tt SGD2PHOT}} $ = 0$; 2) the size of a source need to be larger than the average FWHM of PSF times an empirical factor. While in our selection, as shown above, we use \mbox{{\tt FLUX\_RADIUS\_R}} and \mbox{{\tt CLASS\_STAR}}. In this study, we consider both the large sample (28,815 LRGs) and the one removing the masked images (23,067 LRGs). To distinguish the two, hereafter, we refer the later as the `refined KiDS test sample'. 

\subsection{Strong Lensing Candidates}
\label{sec:4.1}
We first apply our Network to the 28,815 LRGs sample. Adopting the threshold 0.999999 for the final output, our Network identified 8,143 strong lensing candidates. We label them as `machine-candidates'. To further scrutinise the candidates, we conduct human-inspection processes. For that, we prepare a `training set' to train the inspectors. In this set, the positive sample includes real strong lensing systems from HST observations, and the ones in our Network training set. 
The negative sample is the same as that used in the Network training. For the KiDS machine-candidates to be examined, we prepare both the stacked RGB images generated by the STIFF software \citep{Bertin2012}, and the $r$-band images. In addition, we also include the 11 candidates rated high in \cite{Petrillo2017} but missed by our Network into the data set. After being trained, six inspectors independently checked the 8,143+11 images and handed in their rating scores, which are defined as follows

\begin{center}
\emph{definite lens, 2 points;}\\
\emph{{possible lens}, 1 points;}\\
\emph{{non-lens}, 0 points.}\\
\end{center}

\begin{figure}
    \centering
    \includegraphics[scale=0.5]{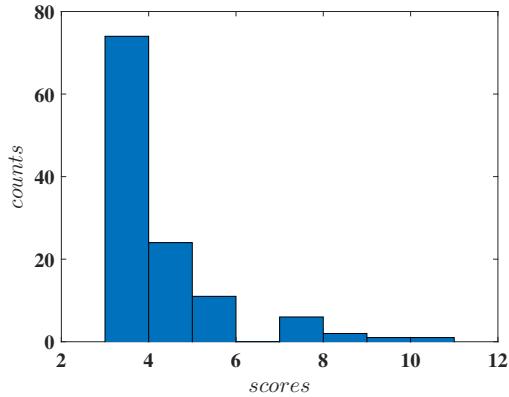}
    \caption{The distribution of scores that the machine-candidates get in human rating. Only those with more than three points are shown here.}
    \label{fig:scores}
\end{figure}

\begin{figure*}
\addtocounter{subfigure}{2}
\centering
    \begin{minipage}{0.2\linewidth}
    \centerline{\includegraphics[width=1\textwidth]{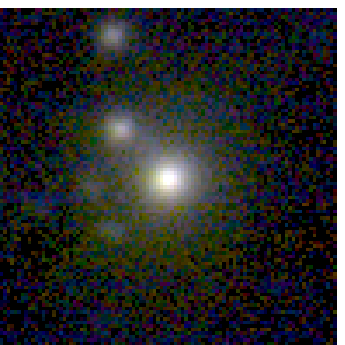}}
    \centerline{J091423.87+022824.74 (4)}
    \hspace{0.15in}
    \end{minipage}
\qquad
    \begin{minipage}{0.2\linewidth}
    \centerline{\includegraphics[width=1\textwidth]{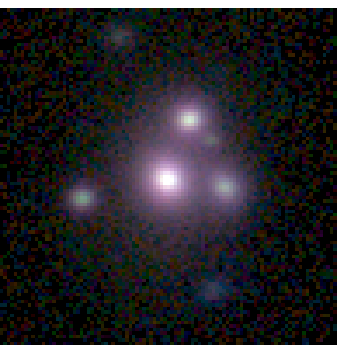}}
    \centerline{J141243.06+004151.44 (4)}
    \hspace{0.15in}
    \end{minipage}
\qquad
    \begin{minipage}{0.2\linewidth}
    \centerline{\includegraphics[width=1\textwidth]{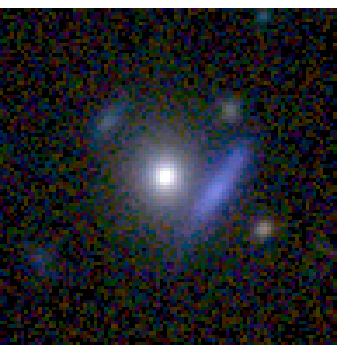}}
    \centerline{J143853.44+003009.15 (4)}
    \hspace{0.15in}
    \end{minipage}
\qquad
    \begin{minipage}{0.2\linewidth}
    \centerline{\includegraphics[width=1\textwidth]{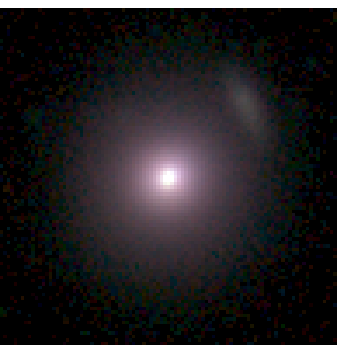}}
    \centerline{J225719.49-304433.13 (4)}
    \hspace{0.15in}
    \end{minipage}
\qquad
    \begin{minipage}{0.2\linewidth}
    \centerline{\includegraphics[width=1\textwidth]{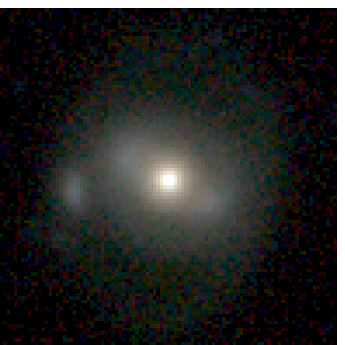}}
    \centerline{J024817.18-320739.71 (3)}
    \hspace{0.15in}
    \end{minipage}
\qquad
    \begin{minipage}{0.2\linewidth}
    \centerline{\includegraphics[width=1\textwidth]{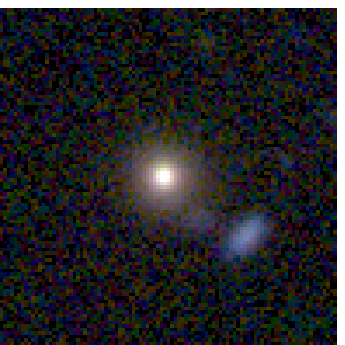}}
    \centerline{J114130.79+003926.05 (3)}
    \hspace{0.15in}
    \end{minipage}
\qquad
    \begin{minipage}{0.2\linewidth}
    \centerline{\includegraphics[width=1\textwidth]{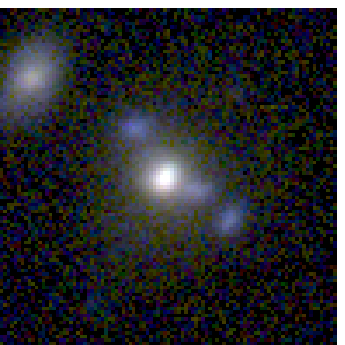}}
    \centerline{J233106.96-340926.62 (3)}
    \hspace{0.15in}
    \end{minipage}

\caption{The 27 new candidates of strong lensing systems that are selected after our human-inspection. They are arranged according to the received scores, from highest to lowest. The source name is given at the bottom of each panel, and the number in bracket is the score received after human-inspection.}
\label{fig:candidates} 
\end{figure*}

We then calculate the total score for each source by summing up the 6 scores from the inspectors. Among the 8,143 images, 43 receive scores of 4 or above; 451 get scores of 3; 600 get scores of 2, and 1,004 receive scores of 1, and all the others have scores of 0. In Figure\,\ref{fig:scores}, we show the score distribution for the 43+451 candidates with scores of 3 or above. With a further voting, we finally select 48 as the high probability strong lensing candidates, 43 with scores of 4 or above, and 5 with scores of 3. We compare them with the ones given by \cite{Petrillo2017, Petrillo2019}, and find that 27 of the 48 candidates are newly identified. We show them in Figure\,\ref{fig:candidates}. Further details of the comparison will be discussed in Section\,\ref{sec:4.1.1-comp-with-petrillo}. 

\begin{figure*}
\centering
    \begin{minipage}{0.2\linewidth}
    \centerline{\includegraphics[width=1\textwidth]{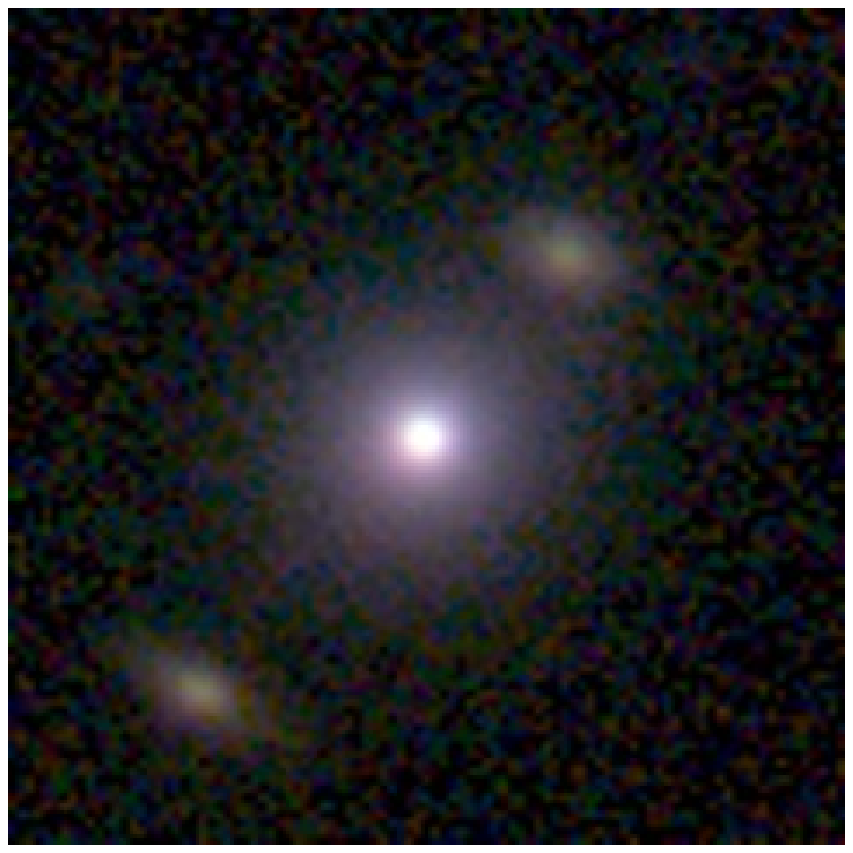}}
    \centerline{J024908.16-305942.40 (8)}
    \hspace{0.15in}
    \end{minipage}
\qquad
    \begin{minipage}{0.2\linewidth}
    \centerline{\includegraphics[width=1\textwidth]{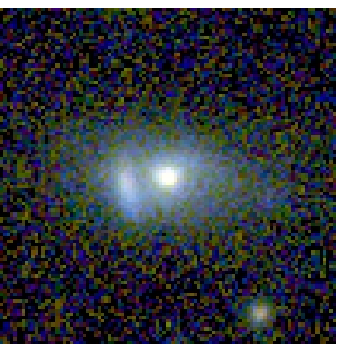}}
    \centerline{J024350.20-311620.40 (7)}
    \hspace{0.15in}
    \end{minipage}
\qquad
    \begin{minipage}{0.2\linewidth}
    \centerline{\includegraphics[width=1\textwidth]{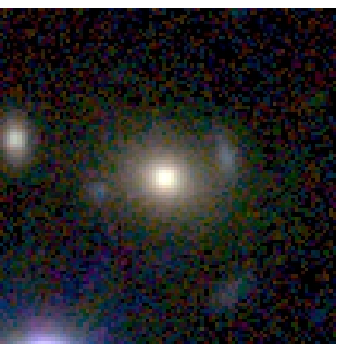}}
    \centerline{J091113.50-000714.17 (7)}
    \hspace{0.15in}
    \end{minipage}
\qquad
    \begin{minipage}{0.2\linewidth}
    \centerline{\includegraphics[width=1\textwidth]{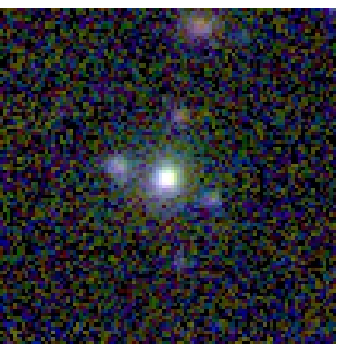}}
    \centerline{J000601.05-323751.42 (5)}
    \hspace{0.15in}
    \end{minipage}
\qquad
    \begin{minipage}{0.2\linewidth}
    \centerline{\includegraphics[width=1\textwidth]{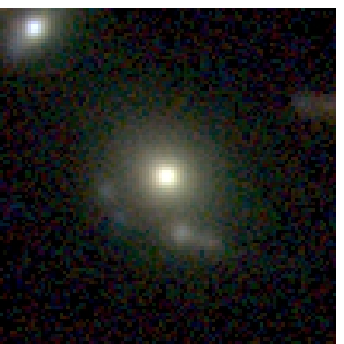}}
    \centerline{J021655.52-335715.50 (5)}
    \hspace{0.15in}
    \end{minipage}
\qquad
    \begin{minipage}{0.2\linewidth}
    \centerline{\includegraphics[width=1\textwidth]{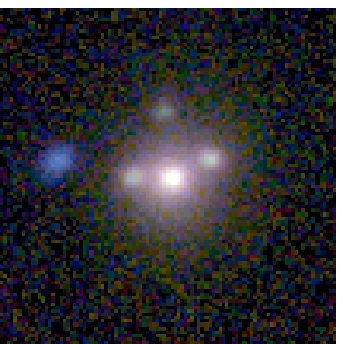}}
    \centerline{J022056.86-330917.46 (5)}
    \hspace{0.15in}
    \end{minipage}
\qquad
    \begin{minipage}{0.2\linewidth}
    \centerline{\includegraphics[width=1\textwidth]{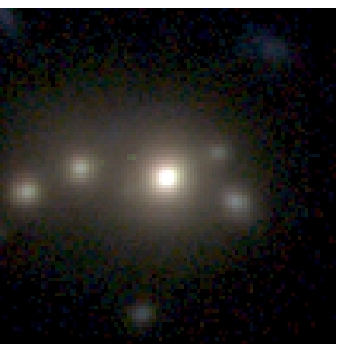}}
    \centerline{J024008.61-330515.71 (5)}
    \hspace{0.15in}
    \end{minipage}
\qquad
    \begin{minipage}{0.2\linewidth}
    \centerline{\includegraphics[width=1\textwidth]{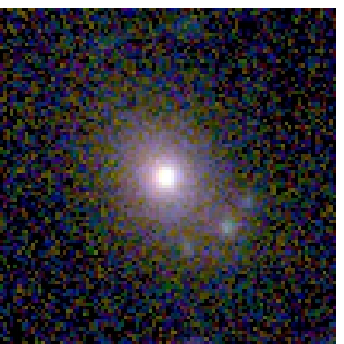}}
    \centerline{J031941.08-334444.68 (5)}
    \hspace{0.15in}
    \end{minipage}
\qquad
    \begin{minipage}{0.2\linewidth}
    \centerline{\includegraphics[width=1\textwidth]{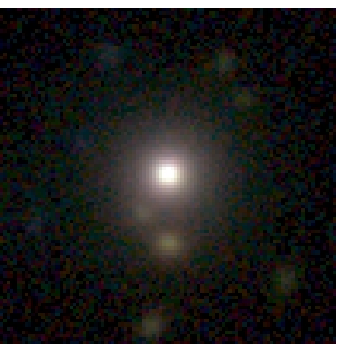}}
    \centerline{J085622.06-013604.23 (5)}
    \hspace{0.15in}
    \end{minipage}
\qquad
    \begin{minipage}{0.2\linewidth}
    \centerline{\includegraphics[width=1\textwidth]{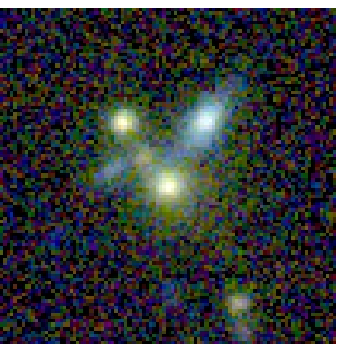}}
    \centerline{J104716.28+005144.19 (5)}
    \hspace{0.15in}
    \end{minipage}
\qquad
    \begin{minipage}{0.2\linewidth}
    \centerline{\includegraphics[width=1\textwidth]{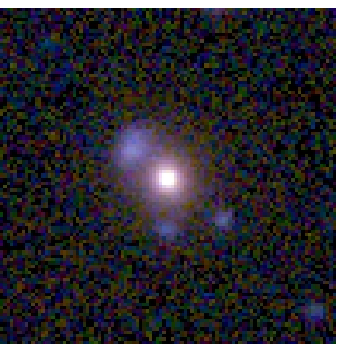}}
    \centerline{J114441.30+003007.63 (5)}
    \hspace{0.15in}
    \end{minipage}
\qquad
    \begin{minipage}{0.2\linewidth}
    \centerline{\includegraphics[width=1\textwidth]{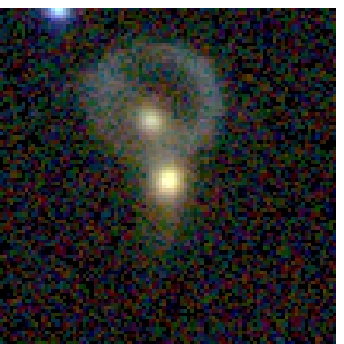}}
    \centerline{J121441.30-020609.86 (5)}
    \hspace{0.15in}
    \end{minipage}
\qquad
    \begin{minipage}{0.2\linewidth}
    \centerline{\includegraphics[width=1\textwidth]{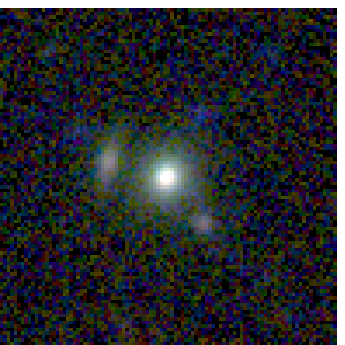}}
    \centerline{J123307.49+000535.40 (5)}
    \hspace{0.15in}
    \end{minipage}
\qquad
    \begin{minipage}{0.2\linewidth}
    \centerline{\includegraphics[width=1\textwidth]{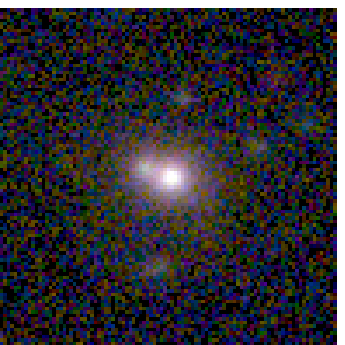}}
    \centerline{J022338.47-331940.53 (4)}
    \hspace{0.15in}
    \end{minipage}
\qquad
    \begin{minipage}{0.2\linewidth}
    \centerline{\includegraphics[width=1\textwidth]{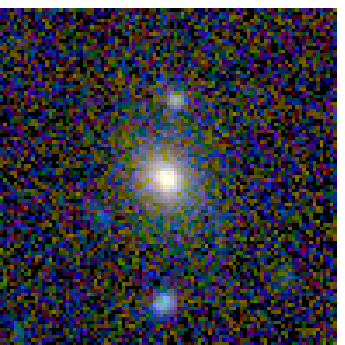}}
    \centerline{J024929.33-310314.59 (4)}
    \hspace{0.15in}
    \end{minipage}
\qquad
    \begin{minipage}{0.2\linewidth}
    \centerline{\includegraphics[width=1\textwidth]{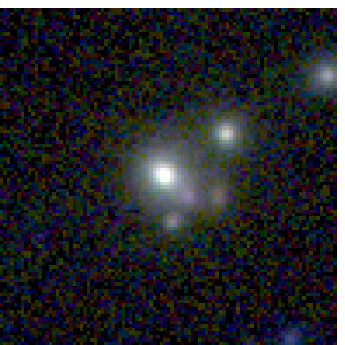}}
    \centerline{J031506.82-322723.49 (4)}
    \hspace{0.15in}
    \end{minipage}
\qquad
    \begin{minipage}{0.2\linewidth}
    \centerline{\includegraphics[width=1\textwidth]{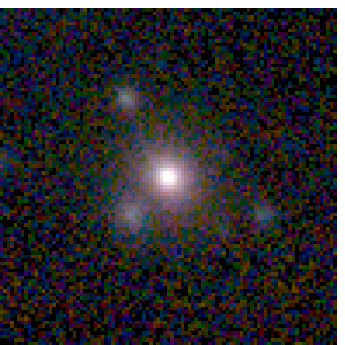}}
    \centerline{J085516.53-012912.44 (4)}
    \hspace{0.15in}
    \end{minipage}
\qquad
    \begin{minipage}{0.2\linewidth}
    \centerline{\includegraphics[width=1\textwidth]{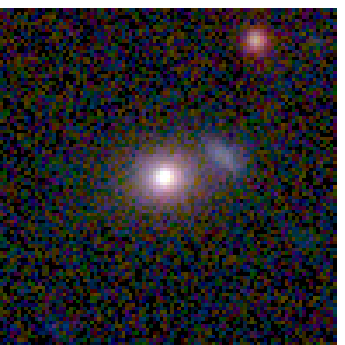}}
    \centerline{J090630.31+005318.88 (4)}
    \hspace{0.15in}
    \end{minipage}
\qquad
    \begin{minipage}{0.2\linewidth}
    \centerline{\includegraphics[width=1\textwidth]{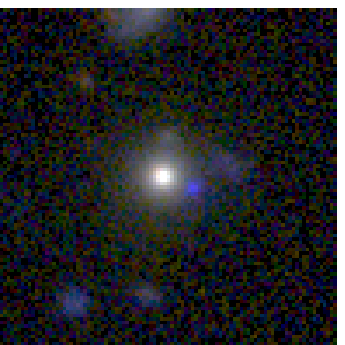}}
    \centerline{J090900.12-014043.80 (4)}
    \hspace{0.15in}
    \end{minipage}
\qquad
    \begin{minipage}{0.2\linewidth}
    \centerline{\includegraphics[width=1\textwidth]{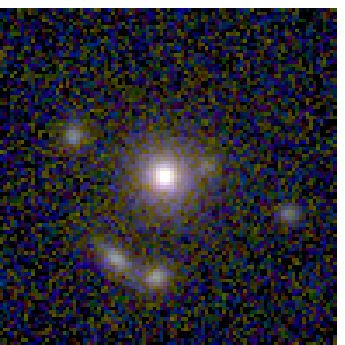}}
    \centerline{J115356.75+001425.97 (4)}
    \hspace{0.15in}
    \end{minipage}
\end{figure*}

\subsection{The training sample with semi-real data}
\label{sec:4.2}
For the training sample applied so far, we construct them fully using simulations. This provides a flexibility to adjust the training data to different survey conditions. However, it may not be able to mimic the specific details of particular observations. The lack of details may lead to high FPR. To study this, we construct another training sample with the methodology similar to that of \cite{Petrillo2017,Petrillo2019}. 

Specifically, to construct the positive sample, we select 2,779 images from the KiDS LRG catalogue presented above as the images of lens galaxies. The following selection criteria are applied: 1) the galaxy images appear to be regular and elliptical; 2) no apparent lensing-like features. For generating lensed images of background sources, we use the same ray-tracing simulations as described in Section\,\ref{sec:training_samples}. The adopted ranges of the parameters are the same with the ones listed in Table\,\ref{parameters, kids, pos}. Similarly as before, the subhaloes are also added here, with the mass function given by Equation\,\ref{eq:sub-massfunc} and with the positions randomly located within the range of $R_{200}$ of the main halo. Again, the light contributions from subhaloes are not considered. A potential shortcoming of this setting for subhaloes is that we may miss the disturbing effect of large subhaloes in the $r$-band images in our training sample. This in turn may affect the performance of our Network. However, we check the 2,779 LRG image stamps, and find that only about 12\% of them contain subclumps identifiable by eyes. Also as we discussed in Section\,\ref{sec:training_samples}, subhaloes are not expected to destroy strong lensing features but only to modify them locally. Thus we do not anticipate a significant impact on our Network performance from ignoring the light contributions of subhaloes to the optical images of LRGs.

Similar as that in \cite{Petrillo2017}, we re-scale the brightness of the lensed background galaxies with respect to that of the foreground LRGs. Specifically,  we checked the typical ratio between the total brightness of the lensed images and the LRGs of our 48 high probability candidates, and find a rescalling range of [0.02, 0.12] to be a suitable one. Then the PSF and the noise are added to the lensed background images. The specific PSF and the noise level are in accord with the corresponding foreground LRG stamp to be superposed on. In the end, we construct 100,000 positive images.

For the negative sample, we also select real LRG images from KiDS data, limiting to those without apparent strong lensing features. 3,037 LRGs are selected including both regular and irregular galaxies. Cases with galaxy pairs and triplets are also included in the negative sample. We randomly rotate these images by 0 to 360 degrees to increase the number of negative images. We also randomly crop out 0 to 5 pixels of the edges. We finally obtain 100,221 non-lens LRGs. Furthermore, we also add 2,020 disk galaxies, which are augmented from 202 disk galaxies from KiDS images through the same augmentation procedure, into the negative sample. In total, we have 100,221+2,020 images in the negative sample. 

\begin{figure*}
\centering
     \begin{minipage}{0.2\linewidth}
     \centerline{\includegraphics[width=1\textwidth, natwidth=95,natheight=95]{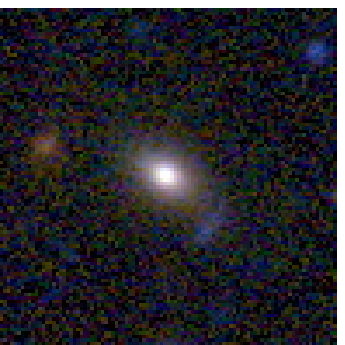}} 
     \centerline{J120626.90+023832.60}
     \hspace{0.15in}
     \end{minipage}
 \qquad
     \begin{minipage}{0.2\linewidth}
     \centerline{\includegraphics[width=1\textwidth, natwidth=95,natheight=95]{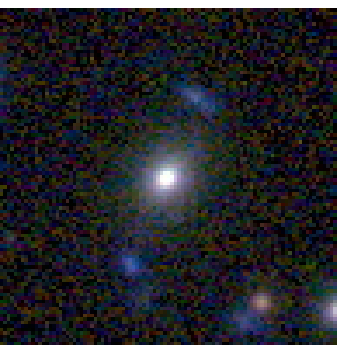}} 
     \centerline{J220050.72-310425.59}
     \hspace{0.15in}
     \end{minipage}
 \qquad
     \begin{minipage}{0.2\linewidth}
     \centerline{\includegraphics[width=1\textwidth, natwidth=95,natheight=95]{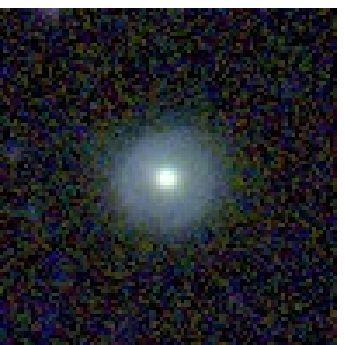}}
     \centerline{J000656.40-333116.25}
     \hspace{0.15in}
     \end{minipage}    
 \qquad
     \begin{minipage}{0.2\linewidth}
     \centerline{\includegraphics[width=1\textwidth, natwidth=95,natheight=95]{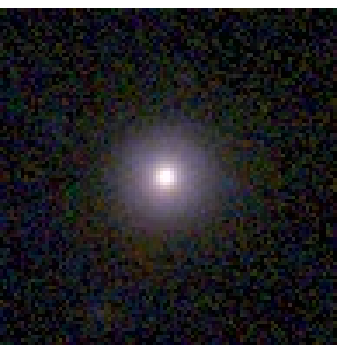}}
     \centerline{J025537.05-312025.16}
     \hspace{0.15in}
     \end{minipage}
 \qquad
     \begin{minipage}{0.2\linewidth}
     \centerline{\includegraphics[width=1\textwidth, natwidth=95,natheight=95]{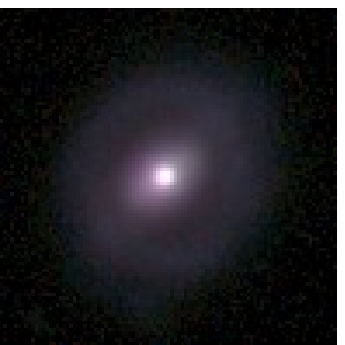}}
     \centerline{J031411.67-301722.39}
     \hspace{0.15in}
     \end{minipage} 
 \qquad
     \begin{minipage}{0.2\linewidth}
     \centerline{\includegraphics[width=1\textwidth, natwidth=95,natheight=95]{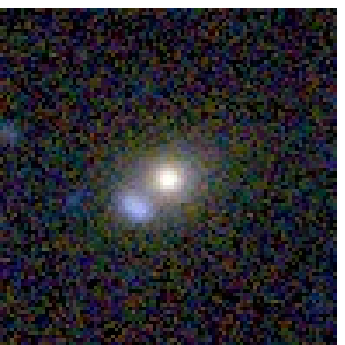}}
     \centerline{J032113.11-303800.06}
     \hspace{0.15in}
     \end{minipage} 
 \qquad
     \begin{minipage}{0.2\linewidth}
     \centerline{\includegraphics[width=1\textwidth, natwidth=95,natheight=95]{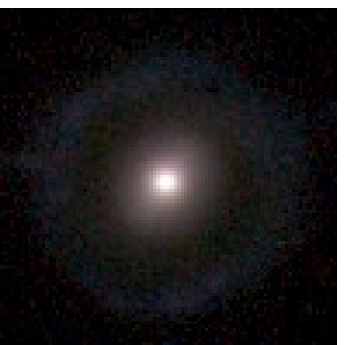}}
     \centerline{J142304.91+001101.23}
     \hspace{0.15in}
     \end{minipage} 
 \qquad
     \begin{minipage}{0.2\linewidth}
     \centerline{\includegraphics[width=1\textwidth, natwidth=95,natheight=95]{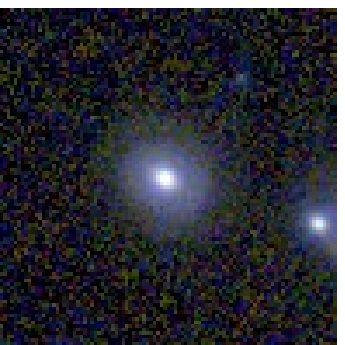}}
     \centerline{J144455.24+000833.08}
     \hspace{0.15in}
     \end{minipage} 
 \qquad
     \begin{minipage}{0.2\linewidth}
     \centerline{\includegraphics[width=1\textwidth, natwidth=95,natheight=95]{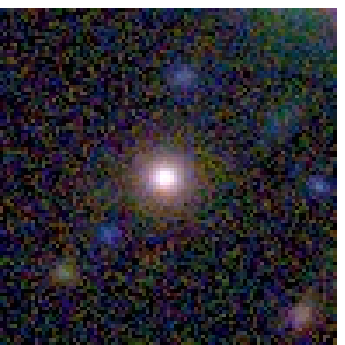}}
     \centerline{J222705.32-320643.63}
     \hspace{0.15in}
     \end{minipage} 
 \qquad
     \begin{minipage}{0.2\linewidth}
     \centerline{\includegraphics[width=1\textwidth, natwidth=95,natheight=95]{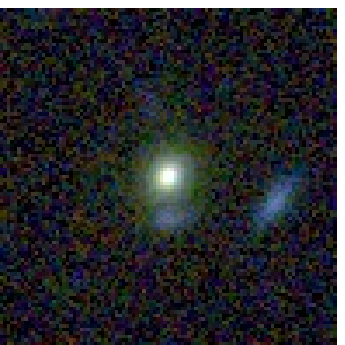}}
     \centerline{J233820.57-323124.88}
     \hspace{0.15in}
     \end{minipage} 
 \qquad
     \begin{minipage}{0.2\linewidth}
     \centerline{\includegraphics[width=1\textwidth, natwidth=95,natheight=95]{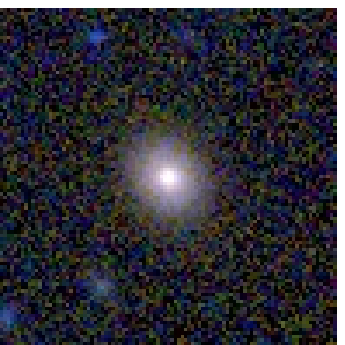}}
     \centerline{J234000.20-325919.68}
     \hspace{0.15in}
     \end{minipage} 
\caption{The 11 supplementary strong lensing candidates that are identified by using the semi-real training set. The name of each candidate is given at the bottom of the stamp.}
\label{fig:candidates-new}
\end{figure*}

To summarise, for the positive sample, we have 100,000 images generated by stacking the real LRG images as the lens and the simulated background lensed galaxy images. The negative sample contains 100,221 non-lens LRGs and 2,020 disk galaxies. After training our Network with this sample, we perform strong lensing candidate searches in KiDS 28,815 LRG test sample. This time, the Network identifies 3,465 candidates. Comparing with the result using the fully simulated training sample, the semi-real training sample can effectively reduce the FPR. We find that 32 of the 48 high probable candidates detected in Section\,\ref{sec:4.1} are also identified here, about a 67\% overlapping fraction. On the other hand, the machine-candidates show some differences due to the different training samples. Among the 3,465 candidates identified here, 1,467 are within the sample of 8,143 discussed in Section\,\ref{sec:4.1}. The rest of the 1,998 candidates are newly found. For them, we visually inspect them by eyes, and classify 17 of them as high probability strong lensing candidates. We put them in our supplement candidate catalogue. Among these 17 candidates, 6 has been identified in \cite{Petrillo2019}. We list them in Table\,\ref{tab:over-lap}, and present the rest of the 11 new candidates in Figure\,\ref{fig:candidates-new}.

\subsection{The Comparisons}
Here we compare the lens-finding results from different aspects. We first compare our results with that of \cite{Petrillo2017,Petrillo2019}. Then we perform internal comparisons between our two results from the fully simulated training sample and from the semi-real training sample respectively. We also discuss the effects of different parameter settings in the training sample.

\subsubsection{Comparison with previous studies}
\label{sec:4.1.1-comp-with-petrillo}
\cite{Petrillo2017,Petrillo2019} have applied their network to KiDS data for detecting strong lensing candidates. In their first study, after human-inspection, they found 56 candidates labelled as the most possible strong lensing systems. Among them, 51 exist in our test sample of 28,815 LRGs. We name them as the comparative sample one, or $CS1$ in short. Within $CS1$, 40 are identified by our Network. In \cite{Petrillo2019}, they provide a list of 1,983 machine-candidates of potential lenses from the LRG sample of KiDS DR4. We refer them as $CS2$, and find 240 of them are picked out by our Network. We also compare our high probability candidates with $CS1$ and $CS2$ samples. We find seven overlapped sources with $CS1$ and 19 overlapped with $CS2$. However, we find that in the 7 and 19 overlapped candidates, 5 of them are in common. Thus, 21 of 48 ($\sim44\%$) of our high probability candidates have been found by previous studies. We list them in Table\,\ref{tab:over-lap}. The rest 27 sources are newly identified by our study. Their 3-band composited images are shown in Figure\,\ref{fig:candidates}. The name and the score from human inspections (in the bracket) are given for each source as well.

It is seen that most of the images in Figure\,\ref{fig:candidates} show features resembling lensing effects, such as J024908.16-305942.40 and J091113.50-000714.17. On the other hand, for J12441.30.020609.86, there is a full ring around one of the two galaxies. It is highly possible that the ring is emerged due to the interaction of these two galaxies, but not a lensing system. We leave it here as an interesting system that deserves further studies. In the 51 candidates of $CS1$, 11 are missed by our Network. Their human-inspection scores are generally lower than 3, with two exceptions receiving scores of 7.

\subsubsection{Internal comparisons}
\label{subsec:4.3.2}
In our studies, we have two different KiDS LRG test samples (KiDS test samples and refined KiDS test samples) as well as two different training samples (the fully simulated one and the semi-real one). It is interesting to compare these results. We label the two Networks trained by the fully simulated training sample and by the semi-real training sample as s-Net and r-Net, respectively. In Table\,\ref{tab:performances-trainingsets-nets}, we list the performance of s-Net and r-Net that are applied to the two KiDS test samples. For convenience, we name the machine-candidates in the case of $a$, $b$, $c$, $d$ as $M1$, $M2$, $M3$ and $M4$ in the table respectively. As expected, $M2$ is a sub-set of $M1$ and $M4$ is a sub-set of $M3$, because the only difference between $case\ a$ and $b$ and between $case\ c$ and $d$ is the testing sample with a further refined one in $b$ and $d$.  
For $M1$ and $M2$, which are the machine-candidates from s-Net, we use 0.999999 as the threshold. For $M3$ and $M4$ from r-Net, we use 0.5 as the threshold. It is noted that in CNN, this threshold is normally very specific to a network architecture, the training sample, and other features. The training sample used in r-Net is more realistic than that of s-Net. We therefore expect a better performance and consequently can lower the threshold comparing to that of s-Net. The value of 0.5 for r-Net is chosen to balance the accuracy and the completeness.  

\begin{figure*}
\centering
    \begin{minipage}{0.2\linewidth}
    \centerline{\includegraphics[width=1\textwidth]{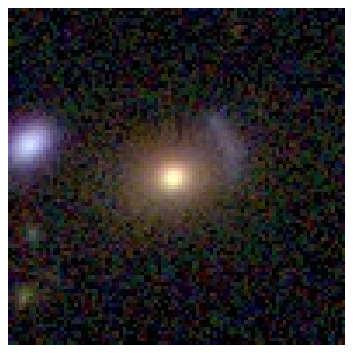}}
    \centerline{J032219.80-342456.55}
    \centerline{\romannumeral1}
    \hspace{0.15in}
    \end{minipage}
\qquad
    \begin{minipage}{0.2\linewidth}
    \centerline{\includegraphics[width=1\textwidth]{figures/J091113_50-000714_17.eps}}
    \centerline{J091113.50-000714.17}
    \centerline{\romannumeral2}
    \hspace{0.15in}
    \end{minipage}
\qquad
    \begin{minipage}{0.2\linewidth}
    \centerline{\includegraphics[width=1\textwidth]{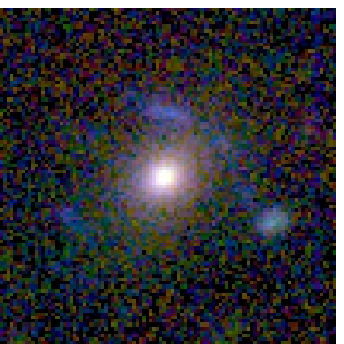}}
    \centerline{J234719.65-333022.44}
    \centerline{\romannumeral3}
    \hspace{0.15in}
    \end{minipage}
\qquad
    \begin{minipage}{0.2\linewidth}
    \centerline{\includegraphics[width=1\textwidth]{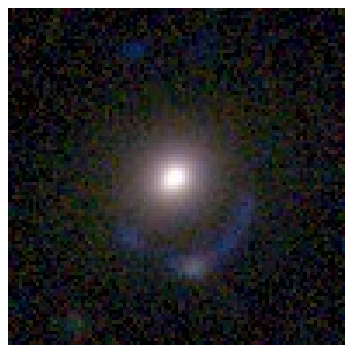}}
    \centerline{J085446.55-012137.14}
    \centerline{\romannumeral4}
    \hspace{0.15in}
    \end{minipage}

\caption{Examples of four strong lensing candidates that are missed when we use different training sets (right two panels) or use different KiDS test samples (left two panels). Stamp \romannumeral1\ and \romannumeral2\ are missed in $case\ b$ but identified in $case\ a$, i.e. they are missed due to using the refined KiDS test sample. Stamp \romannumeral3\ and \romannumeral4\ are found at $case\ a$ but missed at $case\ c\&d$, due to using the semi-real training sample.}
\label{fig:missing-lens} 
\end{figure*}

The FPR is reduced largely from $case\ a$ to $case\ b$. The possible reason is that the masked pixels in the images can form arbitrary shapes. They can mislead the Network, disturb its functionality, and cause a high FPR in $case\ a$. However, it is noted that 10 candidates with high probabilities in $case\ a$ are missed in $case\ b$. Some of them show significant lensing-like features. Two examples are shown in the left two panels of Figure\,\ref{fig:missing-lens}. 

Between $case\ a$ and $case\ c$, there are relatively low overlaps, i.e. only 1,467 candidates appear in both $M1$ and $M3$. Such a difference illustrates the importance of the training sample to the Network. 
The training set composed by semi-real images contains more realistic observational features than the fully simulated one. Thus the FPR is reduced significantly from $case\ a$ to $case\ c$. By further refining the testing sample in $case\ d$, the FPR is further lowered. However, in both $case\ c$ and $case\ d$, some {high probability lensing} candidates are missed comparing to that of $case\ a$ (see \romannumeral3 \romannumeral4 of Figure\,\ref{fig:missing-lens}). 

To summarise, these comparisons show that a more realistic training sample can effectively reduce the FPR. By a refined pre-selection of the data to be analysed, which remove images that may confuse the Network, can further increase the accuracy of strong lensing searches. However, there can be a cost of losing the completeness. Therefore at the moment, it is hard to draw a firm conclusion about the optimal approach. Further investigations are needed about how to perform proper pre-selections in order to balance the accuracy and the completeness.

\begin{table*}
\centering
\begin{tabular}{|c|c|c|c|c|}
\hline
case & LRG samples               & Network & number of lens given by the Network & number of candidates \\ \hline
a    & KiDS test samples         & s-Net   & 8143 ($M1$)                         & 48+0                 \\ \hline
b    & refined KiDS test samples & s-Net   & 4744 ($M2$)                         & 38+0                 \\ \hline
c    & KiDS test samples         & r-Net   & 3465 ($M3$)                         & 32+17                \\ \hline
d    & refined KiDS test samples & r-Net   & 2877 ($M4$)                         & 27+13                \\ \hline
\multicolumn{5}{|c|}{number of overlapping}                                                             \\ \hline
\multicolumn{5}{|c|}{$M1 \cap M2$: 4744; $M1 \cap M3$: 1467; $M1 \cap M4$: 1211;}                       \\
\multicolumn{5}{|c|}{$M2 \cap M3$: 965;  $M2 \cap M4$: 965; $M3 \cap M4$: 2877.}                     \\ \hline
\end{tabular}
\caption{The performance of our Network in different cases. The details of KiDS test samples and refined KiDS test samples are given at the beginning of  Section\,\ref{sec:lensing_candidates}. See the definitions of r-Net and s-Net in Section\,\ref{subsec:4.3.2}. {The numbers in the last column show how many high probability candidates found in $case\ a$ or $b$ are also identified in $case\ c$ or $d$, respectively (first number), and the newly identified ones in $case\ c$ or $d$ (second number).} In the last two rows, the sizes of overlapping of the samples are given.}
\label{tab:performances-trainingsets-nets}
\end{table*}

\subsubsection{The distributions of the model parameters}
\label{subsec: 4.3.3}

For both the fully simulated training sample and the semi-real one, we use flat distributions for galaxy model parameters. Because in observation these parameters show certain distributions rather than flat ones, it is desirable to test if the chosen parameter distributions affect the Network performance significantly. We perform tests to see the effect of the distribution of the Sersic index. For that, we construct 40 training sets from a large parent sample containing 1000,000 fully simulated images. Among them, 20 sets have flat distributions between $n=2$ to $5$ for the Sersic index as before. For the other 20 sets, we choose the training images so that the Sersic index follows the distribution of \cite{Huertas-Company2013} within the range of 2 to 5. Each of the 40 sets contains 100,000 images, 50,000 positive and 50,000 negative. The Network is then trained by the two groups of training sets, separately, and applied to the KiDS LRG test sample. We compare the outputs from the two groups with the training sets having different Sersic index distributions. The overlapping fractions with the 8,143 machine-candidates and with the 48 high probability candidates from s-Net using our fiducial training sample are shown in Table\,\ref{tab:sersic test}. We can see that using training samples with different distributions of the Sersic index, the Network returns similar results. It indicates that the flat distributions adopted in our fiducial training sample shall not affect our Network performance significantly.

\begin{table}
\begin{tabular}{|c|c|c|}
\hline
      & overlap with  & overlap with\\ 
      &8143 machine-candidates   &48 high probability candidates\\
\hline
non-flat  & 77.80\%         & 80.31\%        \\ \hline
flat    & 78.73\%           & 81.25\%        \\ \hline
\end{tabular}
\caption{Impacts of the Sersic index distributions. The first column is the average percentage of overlapping with the 8,143 machine-candidates. The second column is the average percentage of overlapping with the 48 high probability candidates.}
\label{tab:sersic test}
\end{table}

\section{Tests of the robustness}
\label{sec:test_the_network}
{ Although CNN has been widely applied in searching for strong lensing systems \citep[e.g. ][]{Ostrovski2017,Schaefer2018,Ma2018,Petrillo2019}, several issues remain elusive, e.g. the stability and robustness of the network. In this section, we investigate how our trained Network performs if the testing data have a wider range of PSF than that in the training data. We also investigate how the volume of the training sample affects the Network performance. It is noted that the detailed testing results may depend on our specific Network. However, as our Network is in the class of AlexNet, we expect that the trends seen from our tests here can be meaningful to other networks with similar structures.}

\subsection{PSF test}
\label{sec:psf tst}
With current techniques, the performance of a network is highly dependent on its training samples. Thus for a specific survey, it requires to build its own training samples to take into account the typical observational characteristics. However, even within a survey, the observational conditions can vary significantly from time to time. Thus it is interesting and important to test the robustness of a trained Network under varying observational conditions. Here, we carry out studies on how the varying PSF affects our Network's performance in strong lensing searches.

We consider the Network of $case\ a$ in Table\,\ref{tab:performances-trainingsets-nets}. In the training sample, the FWHM of PSF is randomly taken from 0.55 to 0.75 arcsec, a narrow range around the typical KiDS PSF of 0.65 arcsec. To test the performance of the trained Network under different PSFs, we create different validation samples by considering different PSF sizes. Specifically, the PSF sizes of (0.4, 0.6, 0.8, 1.0, 1.2, 1.4, 1.6, 1.8, 2.0) times 0.65 arcsec, are analysed. Each validation set consists of 2,000 lensing images and 2,000 non-lensing images. These images are generated by simulations following the same procedures as those for the training sample except with different PSFs. Totally, we have nine validation sets. We apply our trained Network to them to test the effect of different PSFs on the Network's performance. 

\begin{figure*}
\centering
\centering
\includegraphics[width=3in]{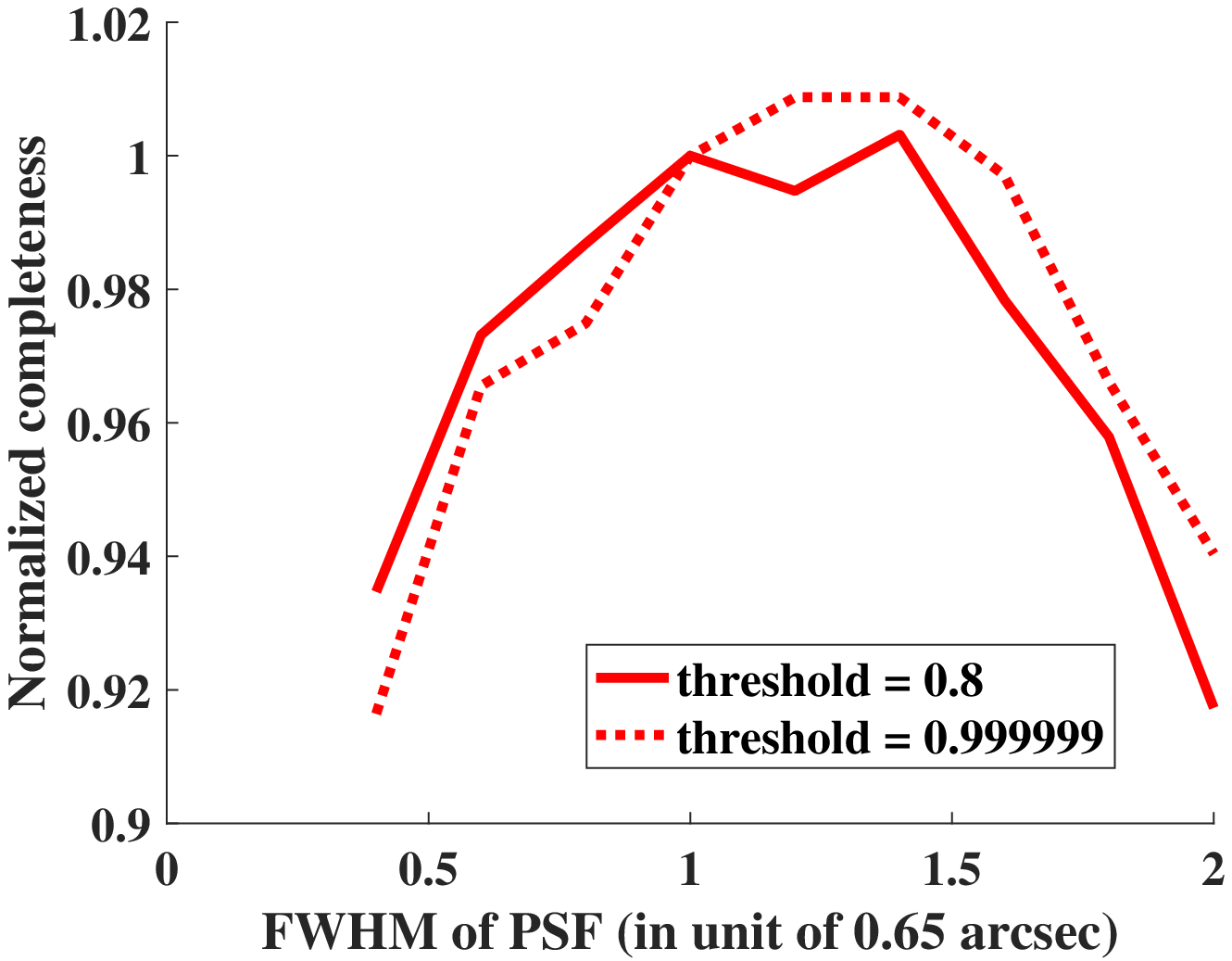}
\includegraphics[width=3in]{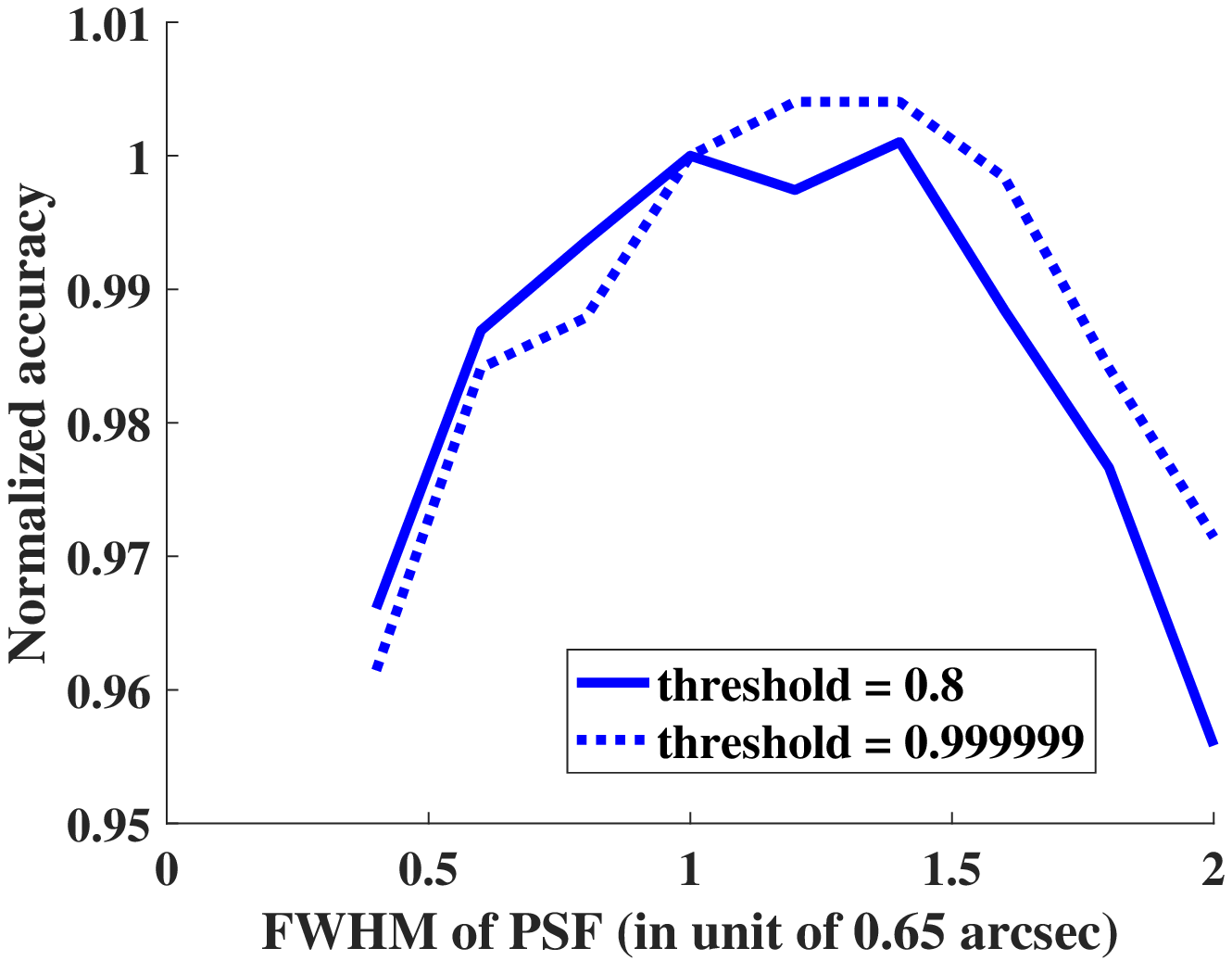}
\centering
\caption{The scores of the Network when we use different PSFs in the validation sets. Different lines show the scores that calculated in different threshold. Left: completeness; right: accuracy.}
\label{fig:psf_result}
\end{figure*}

In Figure\,\ref{fig:psf_result}, we show the completeness (left panel) and the accuracy (right panel) versus the size of PSF. The results are normalised to the ones with PSF of 0.65 arcsec. We consider different output thresholds, which are presented by solid and dashed lines, respectively. Similar trends can be found in both panels, and the best performance is achieved when the PSF is close to that in the training sample. There are degradation toward both larger and smaller sizes of PSF.
However, the overall degradation of the Network's performance is small ($<10\%$). This suggests that the Network trained using a narrow PSF range around the median PSF can be safely applied to data with a relatively broader PSF variations.

\subsection{The volume of the training sample}
\label{sec:size tst}
For a network, it usually requires a training set with a large enough volume and being a fair representation of the data set to be investigated. There are however, no quantitative studies yet on exactly how large a training set is sufficient. It depends on the learning ability of a network and the problem to be investigated. For strong lensing searches, in literature, different sizes of training samples have been used, from tens of thousands galaxies \citep{Ma2018} to a million \citep{Petrillo2017}. Here we perform quantitative analyses on how the size of the training sample affects the performance of our Network. 

\subsubsection{Performance test}
\label{sec:5.2.1}
To test how the size of the training sample affects the performance of our Network, we construct a large training set containing one million images with KiDS conditions as described in Section\,\ref{sec:training_samples}. This is also the parent sample used in Section\,\ref{subsec: 4.3.3}. From it, we then build different training samples with different sizes. For validation, we generate four data sets by simulation, each consisting of 2,000 images with half lensing systems and half negative images, but with different random seeds. 

\begin{figure}
    \centering
    \includegraphics[scale=0.5]{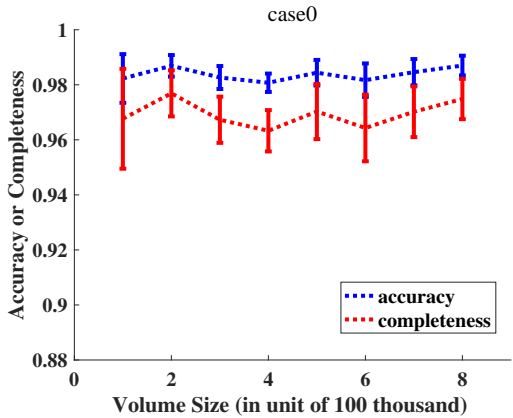}
    \caption{The accuracy and completeness in the volume size test.
    Default parameters of Network is used, i.e. $case\ 0$. (Section\,\ref{sec:5.2.1}).}
    \label{fig:size_result_case0}
\end{figure}

From the one million parent training images, we randomly select different subsets to create training samples with different sizes. The sizes are: 0.1, 0.2, 0.3, 0.4, 0.5, 0.6, 0.7, 0.8 times a million, respectively. For each size, we generate five different training samples from the parent images, and train the Network accordingly. 
It is noted that in our default setting, we adopt the network parameters with the learning rate $\eta$=0.03, the batch size {\it bsize}=32 and {\it epochs}=50. We denote this setting as $case\ 0$. We then apply the trained Network to the validation sets. Adopting the threshold of 0.8, the results are shown in Figure\,\ref{fig:size_result_case0}. The data points are the average over 20 sets, i.e. at each volume size, five trained Networks are applied to four validation sets respectively. The error bar is the standard derivation of the 20 sets. It is seen that the result does not vary significantly with the increase of the volume size. For the completeness, the difference between the best and the worst is only about 0.6\%. For the accuracy, it is about 1\%. Therefore, our tests suggest that $\sim 0.2$ million images in the training sample should be sufficient for our Network to reach a stable performance.

\begin{figure*}
\begin{minipage}[t]{0.49\linewidth}
\centering
\includegraphics[width=3.0in]{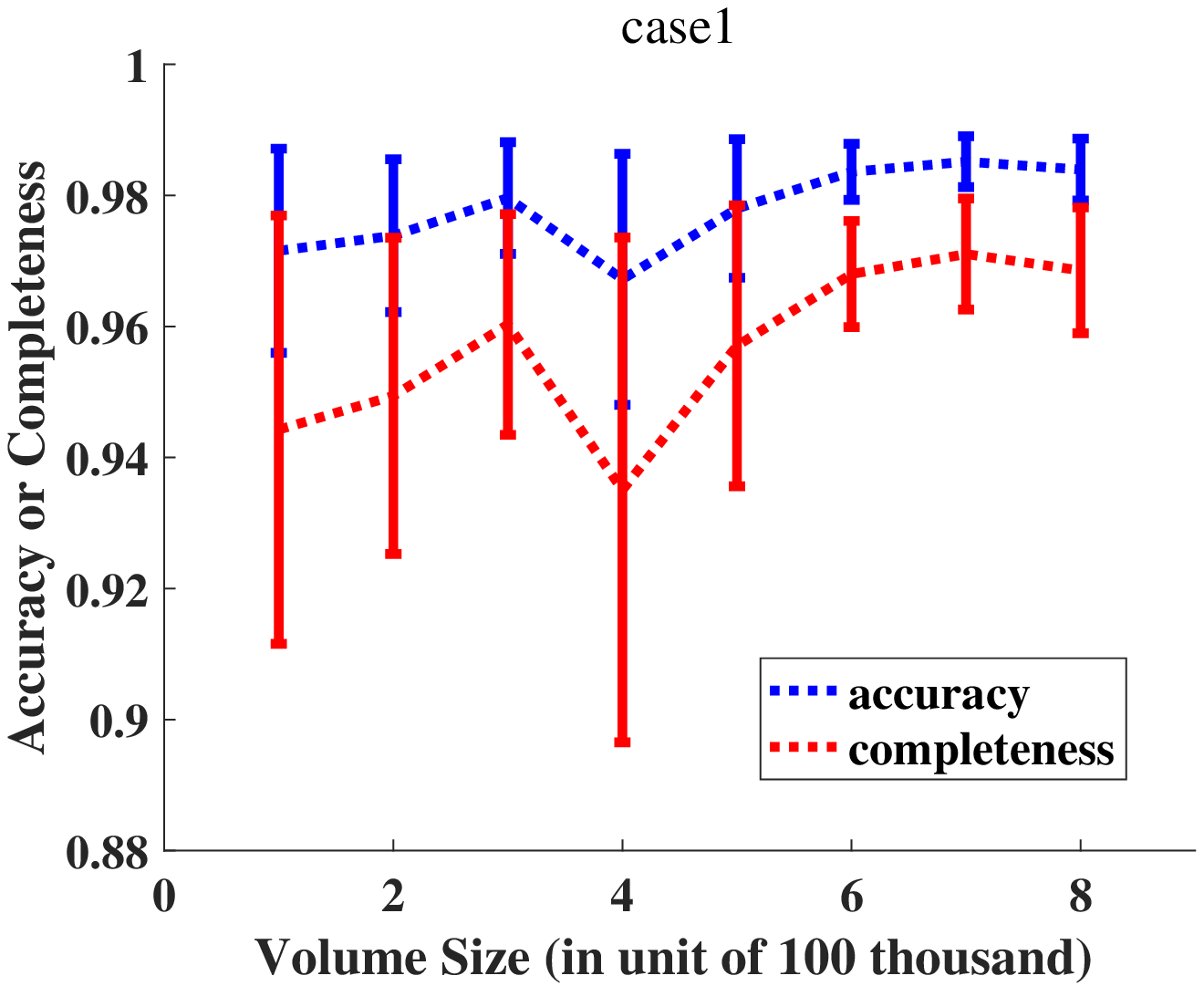}
\includegraphics[width=3.0in]{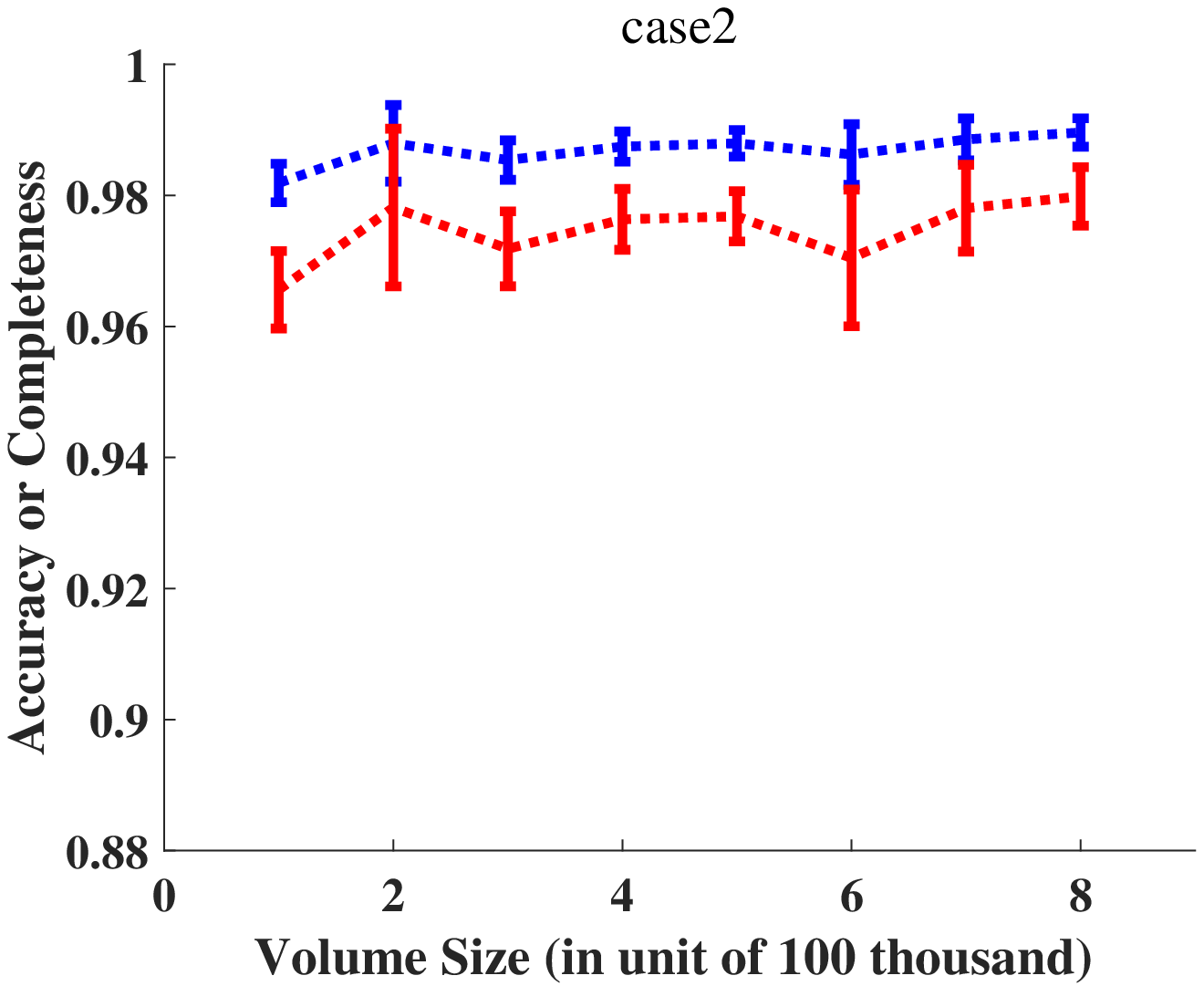}
\end{minipage}
\begin{minipage}[t]{0.49\linewidth}
\centering
\includegraphics[width=3.0in]{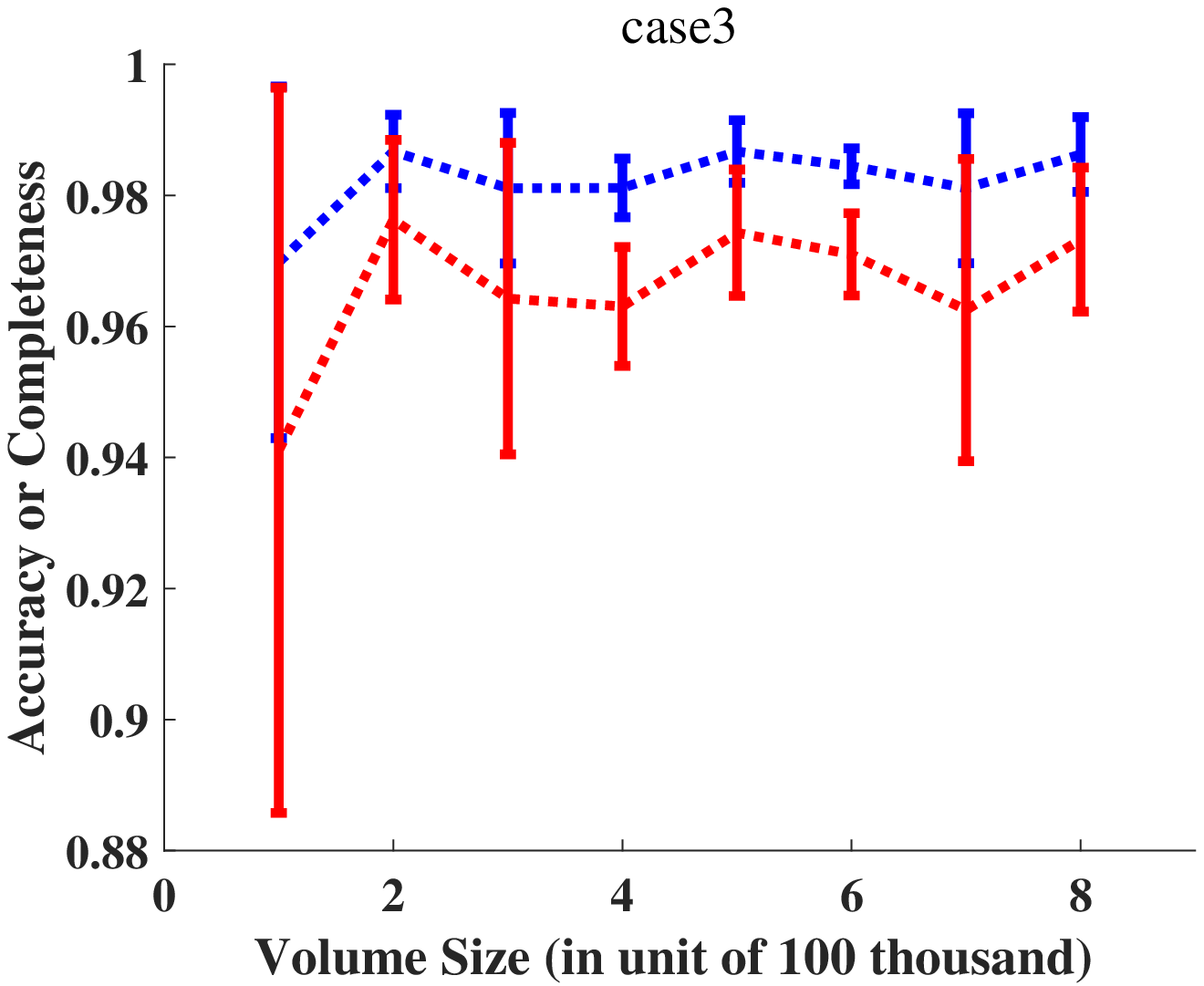}
\includegraphics[width=3.0in]{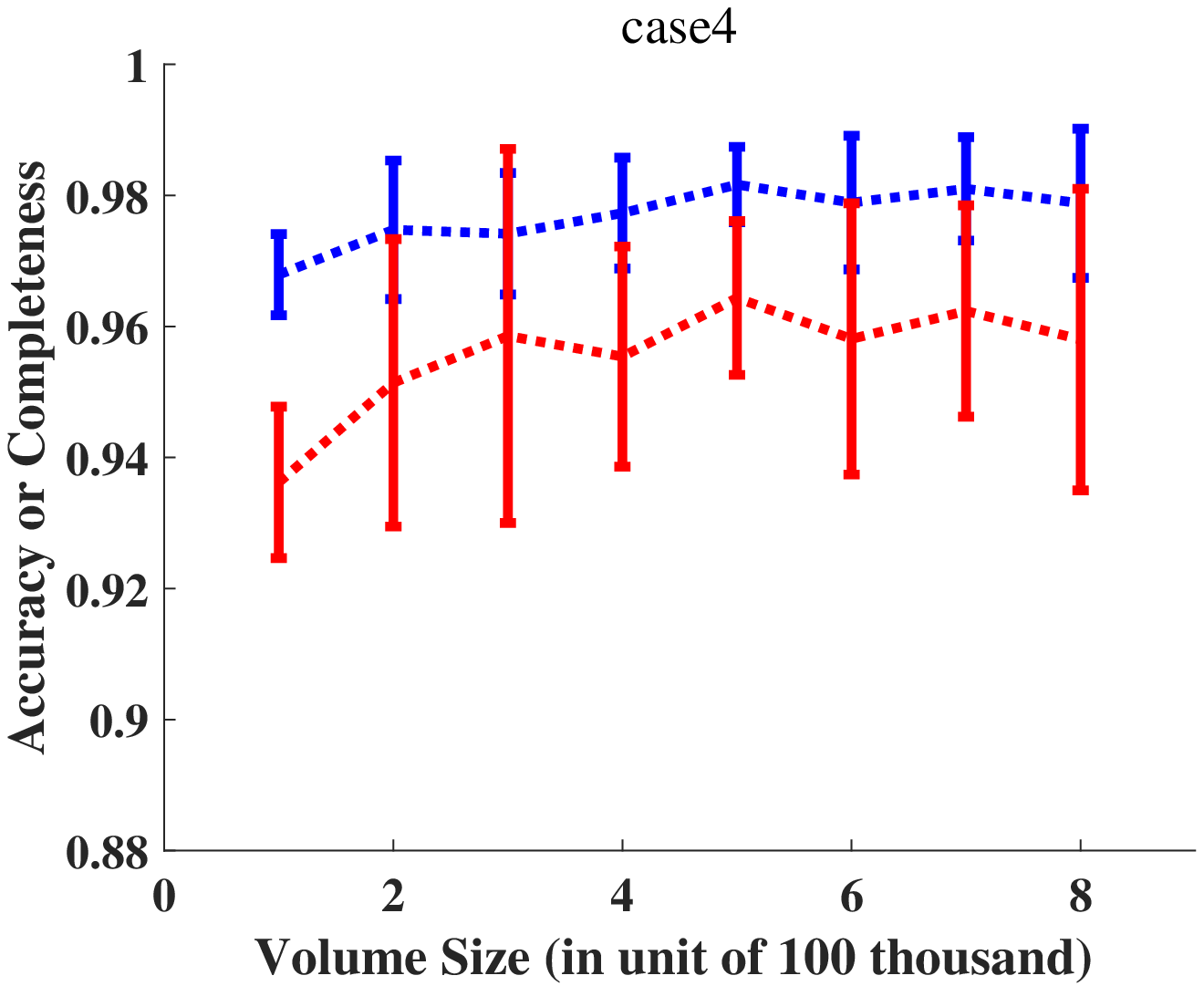}
\end{minipage}
\caption{Same as Figure\ref{fig:size_result_case0} but with different Network parameters. The upper left is the performances when the $\emph{$\eta$}$ is set from 0.03 to 0.01, i.e. $case\ 1$; the bottom left is the performances when the $\emph{epochs}$ is set from 50 to 100, i.e. $case\ 2$; the upper right is the performances when the $\emph{bsize}$ is set from 32 to 64, i.e. $case\ 3$; the bottom right is the performances when the $\emph{$\eta$}$ is set from 0.03 to 0.01 and the $\emph{bsize}$ is set from 32 to 64, i.e. $case\ 4$.}
\label{fig:size_result_hst_changing_para}
\end{figure*}

We also test the effect of the volume size under different Network parameters. In $case\ 1$, we change the learning rate to be $\eta$=0.01 and the other parameters are the same as those in $case\ 0$. In $case\ 2$, we change the {\it epochs} to be 100. In $case\ 3$, {\it bsize} is set to be 64. Finally in $case\ 4$, $\eta$=0.01 and {\it bsize}=64. The testing results are shown in Figure\,\ref{fig:size_result_hst_changing_para}. Again, we see a rather stable performance in different cases showing the insensitivity of the trend to the network parameters. In Table\,\ref{tab:performance_case_one_to_four}, the average performance and the standard deviations are shown. We can see that the average performances are about the same in different cases. On the other hand, the standard variations show some differences. This implies that the network parameters can influence the result from a single training and a validation. The optimal training size for each case also varies somewhat although different lines in Figure\,\ref{fig:size_result_hst_changing_para} are rather flat. For all the considered cases, conservatively, $\sim 0.6$ million is sufficient to train our Network.

\begin{figure*}
    \centering
    \includegraphics[width=3.0in]{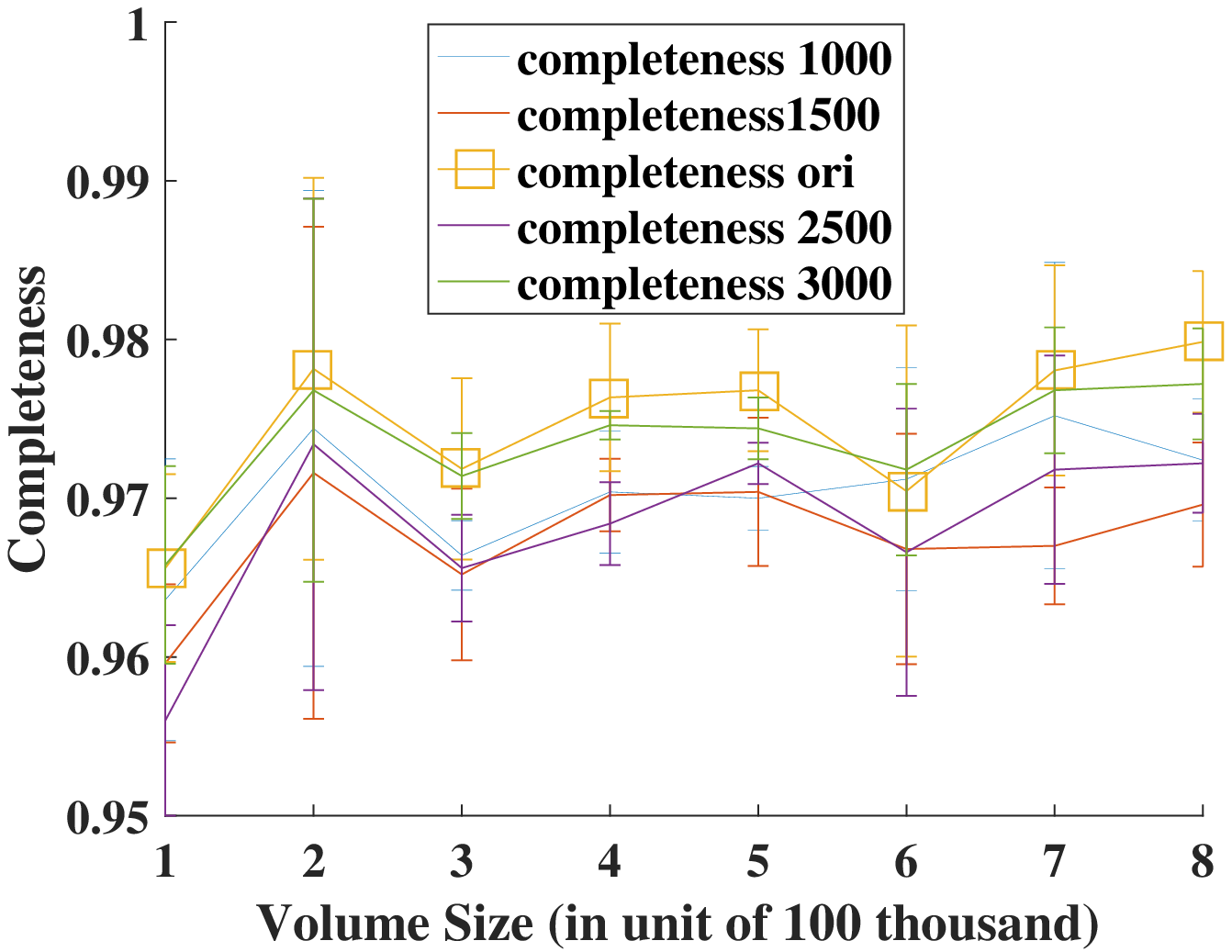}
    \includegraphics[width=3.0in]{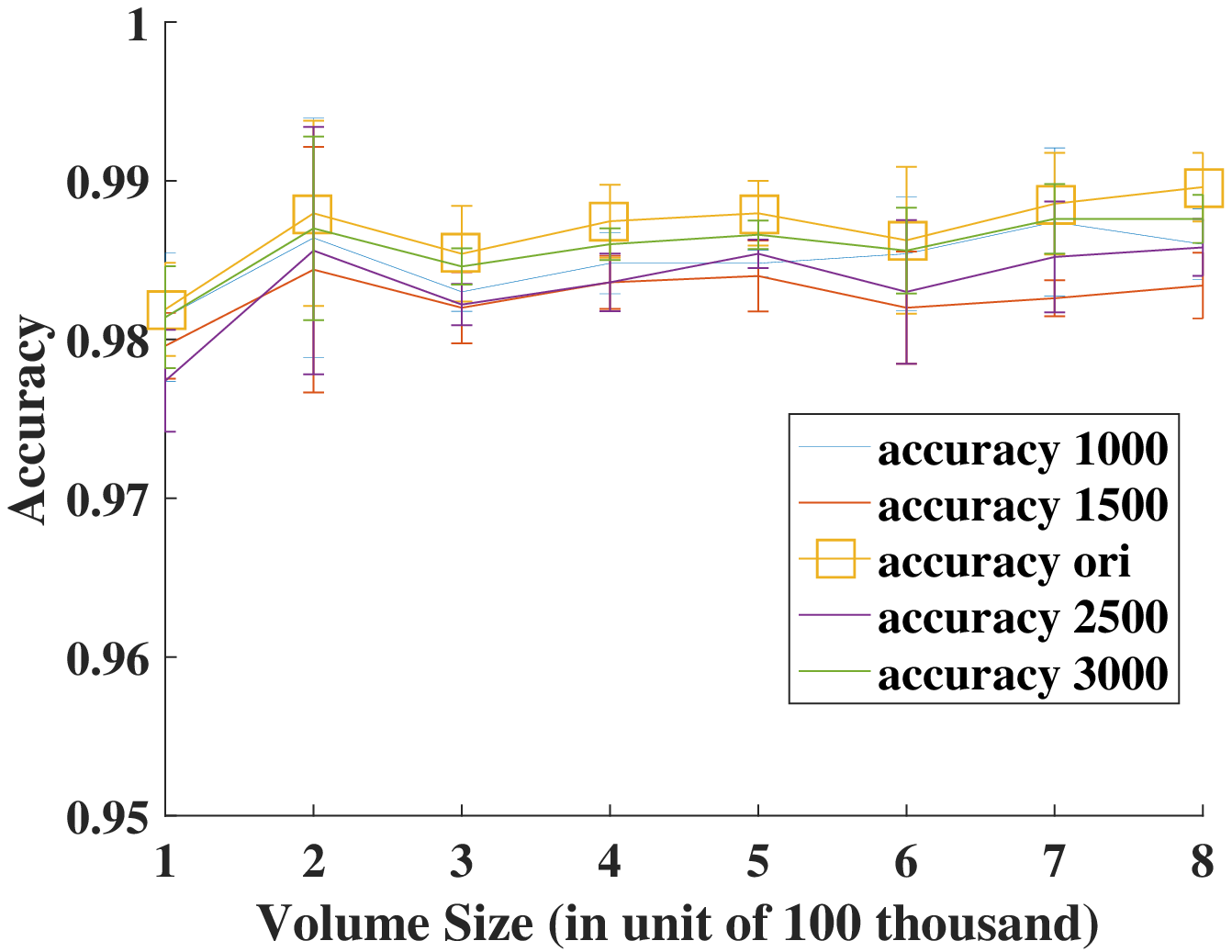}
    \caption{Same as Figure \ref{fig:size_result_hst_changing_para} but with different validation sets. Network parameters of $case\ 2$ in Table\ref{tab:performance_case_one_to_four} ($\emph{$\eta=0.03$}$, $\emph{$bsize=32$}$ and $\emph{$epochs=100$}$) are used. The curves of `ori' are calculated based on the original simulated validation sets, i.e. the ones used for Figure\,\ref{fig:size_result_case0}. The others are calculated using new validation sets, with the size shown by the number.}
    \label{fig:retest_of_lr_bsize}
\end{figure*}

To see if the results depend on the size of the validation set, we  generate 4 more validation sets with different sizes. The $case\ 2$ parameters in Table\,\ref{tab:performance_case_one_to_four} are used, and the results are shown in Figure\,\ref{fig:retest_of_lr_bsize}. It is seen that the results are not sensitive to the size of the validation set. We also analyse the results with different output thresholds, and the differences are minimal.


\begin{table}
\begin{tabular}{|c|cc|cc|}
\hline
      & \multicolumn{2}{c|}{scores} & \multicolumn{2}{c|}{error bar($\times 10^{-4}$)} \\ \cline{2-5} 
      & accuracy   & completeness   & accuracy              & completeness             \\ \hline
$case\ 0$ & 0.984      & 0.969          & 0.492                 & 1.01                     \\ \hline
$case\ 1$ & 0.978      & 0.957          & 0.978                 & 2.00                     \\ \hline
$case\ 2$ & 0.987      & 0.975          & 0.327                 & 0.671                    \\ \hline
$case\ 3$ & 0.982      & 0.966          & 0.913                 & 1.88                     \\ \hline
$case\ 4$ & 0.977      & 0.956          & 0.872                 & 1.88                     \\ \hline
\end{tabular}
    \caption{Left two columns and right two columns are the average scores and standard deviations respectively. They are calculated based on all the training samples with different sizes in the corresponding cases. }
    \label{tab:performance_case_one_to_four}
\end{table}

\section{Summary and discussion}
\label{sec:summary}

\subsection{Main conclusions}
We construct a CNN network to search for galactic-scale strong lensing systems. Two approaches are adopted in preparing the training samples. One is to generate the lensing and non-lensing images fully by simulations. The other is to combine real observational images of foreground LRGs with simulated background sources. The Network is then trained by the two sets of samples, respectively. Applying the Network trained by the fully simulated sample to KiDS DR3 LRG data, we find about 8,000 machine-candidates of strong lensing systems. After human inspection, we identify 48 high probability candidates. Among them, 21 are overlapped with those in \cite{Petrillo2017,Petrillo2019}, and 27 are newly found ones. Using the Network trained by the semi-real sample, we find 17 additional high-probability candidates. Among them, 6 are overlapped with those in \cite{Petrillo2019}, and the rest 11 are newly identified.

The difference between the results from our studies and the ones from \cite{Petrillo2017} may be attributed to the different network structures and the training and test samples. While both are AlexNet based, our Network contains two convolutional layers and three full connection layers, similar to that of \cite{Davies2019}. On the other hand, the network used in \cite{Petrillo2017} has four convolutional layers and two full connection layers. For the training samples, when we use the fully simulated one, 8,143 machine-candidates are located. With the semi-real training sample, the number decreases to 3,465. Further refining the KiDS LRG test sample by excluding images containing apparent masks reduces the number of machine-candidates to 2,877. The last number is best comparable with the results of \cite{Petrillo2017} with similar data settings, in which they found 761 machine-candidates.
We need to point out that the different criteria in the pre-selection of the LRG sample also contribute to the differences between our results and that from \cite{Petrillo2017,Petrillo2019}. Our refined test sample has 23,067 LRGs, while theirs is 21,798. 

It is worth mentioning that while using the semi-real training sample and the refined test sample can effectively reduce the FPR, there are some cost of completeness. Further investigations are needed to find optimal designs for the training sample and the proper pre-selections for the test data. 

In this paper, we also carry out, for the first time, systematic analyses about the dependence of the Network performance on the PSF of the test data, and on the size of the training sample. For the former, considering different test data with PSF varying from 0.4 to 2 times of the median PSF that is used to train the Network, the overall performance degradation is less than 8\%. This shows the feasibility of using our Network trained with a narrow range of PSFs to search for strong lensing candidates from data with a relatively broader PSF variations.

For the dependence on the training volume, we construct different training samples with volume ranging from 0.1 to 0.8 millions. Several Network parameters are also varied to see the robustness of the results, including the learning rate, the batch size and the number of epochs. We find that within the considered range, the performance of our Network is not sensitive to the volume size, and a sample consist of $\sim 0.2$ million images is sufficient to train our Network. Fluctuations exist between individual runs with the level depending somewhat on the Network parameters. Our results here is instructive to other networks with similar architectures. For more complex networks with more parameters, larger training samples may be needed.

In addition to the volume of the training sample, the characteristics of the galaxies included in the training data also affect the Network performance. This is particularly important for the negative sample. For that, we perform a series of tests to include different types of galaxies in the negative sample. The negative sample containing only single elliptical galaxies leads to low efficiency.
With the fraction of 80\%, 10\% and 10\% of elliptical (including both single and paired ones), highly elongated and spiral galaxies in the negative sample, our Network reaches the optimal performance. 

\subsection{Limitations and future improvements}
\label{sec:future plans}
There are several aspects that we can further improve in future studies. First, we only use $r$-band images in searching for strong-lensing candidates, i.e. only the morphological information is used. It is known that adding in colour information of galaxies can improve the strong-lensing search efficiency significantly \citep{Metcalf2019}. One of our future tasks is to extend our Network to include the images from multiple bands.

In our image simulations, the adopted galaxy parameters incline to use elliptical galaxies. While they are suitable for LRG lens galaxies, the other types of source galaxies are missing in our study. We argue that for single $r$-band images, this shortage will not affect the performance dramatically. However, when we incorporate multi-bands information in future, such simplicity will be critical. In our future studies, we plan to include background galaxies with diverse types, and correspondingly to adopt suitable parameter ranges and distributions.

For subhaloes, in our current setting, we consider the subhalo contributions to the lensing effect, but not to the light distributions of images. While this is tolerable in our analyses with KiDS data, future high resolution and high sensitivity observations will demand more sophisticated modelling of subhaloes. Moreover, the line-of-sight structures should also be taken into account carefully. 

In this study, we adopt an AlexNet-based network with two convolutional layers and three full connection layers. This is similar to AstrOmatic, GAMOCLASS and NeuralNet2 \citep{Davies2019} presented in Strong Lens Finding Challenge \citep{Metcalf2019}. For other network structures, our conclusions about the PSF variation and the volume of the training sample may not hold. As one of our future efforts, we plan to implement more network structures in order to understand different issues more comprehensively.

\section*{Acknowledgements}
We thank the anonymous referee for very valuable and constructive comments that help us to improve the paper significantly.
We thank Yiping Shu, Dandan Xu, Giovanni Covonne (SWIFAR visiting fellow), Rui Li, Nan Li, Nicola R. Napolitano and Toshi Futamase (SWIFAR visiting fellow) for the extensive discussions.
This work is supported by NSFC Grant No. 11933002, 11773074 and 11803028, and by YNU Grant C76220100008. Z.H.F. and X.K.L. are also supported by a grant of CAS Interdisciplinary Innovation Team.

\section*{Data Availability}
The KiDS DR3 data used in our study are fully public and can be accessed via the link http://kids.strw.leidenuniv.nl/DR3/index.php. Our strong lensing machine candidates of $M1$ and the ones listed in Table\,\ref{tab:over-lap} are available in Github at https://github.com/EigenHermit/show\_stamp/.

\bibliographystyle{mnras}
\bibliography{citationlist.bib}

\appendix
\section{The redshift combinations}
In Table\,\ref{redshift combination}, we list the redshift combinations that are used for the simulations in Section\,\ref{sec:training_samples}.
\begin{table}
\centering
\begin{tabular}{c|cccc}
\hline
$z_d$ & \multicolumn{4}{c}{$z_s$} \\ \hline
0.01  & 0.02  & 0.03 & 0.06 & 3.5 \\ \hline
0.1   & 0.2   & 0.3  & 0.5  & 3.5 \\ \hline
0.2   & 0.4   & 0.55 & 0.9  & 3.5 \\ \hline
0.3   & 0.5   & 0.7  & 1.15 & 3.5 \\ \hline
0.4   & 0.6   & 0.85 & 1.35 & 3.5 \\ \hline
0.5   & 0.8   & 0.9  & 1.05 & 3.5 \\ \hline
0.6   & 0.9   & 1.2  & 1.8  & 3.5 \\ \hline
0.7   & 1.1   & 1.4  & 2.0  & 3.5 \\ \hline
0.8   & 1.3   & 1.65 & 2.2  & 3.5 \\ \hline
0.9   & 1.4   & 1.7  & 2.3  & 3.5 \\ \hline
1.0   & 1.6   & 1.95 & 2.5  & 3.5 \\ \hline
\end{tabular}
\caption{The redshift combinations that are used in our simulations.}
\label{redshift combination}
\end{table}

\section{Overlapped sources}
We list the overlapped lensing candidates with \cite{Petrillo2017,Petrillo2019} among our high probability candidates and the supplementary candidates in Table\,\ref{tab:over-lap}.\footnote{The corresponding RGB and fits images can be found from https://github.com/EigenHermit/show\_stamp. The images of the whole machine candidates of $M1$ are also included.}

\begin{table}
\centering
\begin{tabular}{c}
\hline
source name                                 \\ \hline
overlapped with high probability candidates \\
KIDS J032219.80-342456.55                   \\
KIDS J234719.65-333022.44                   \\
KIDS J114612.74-024632.00                   \\
KIDS J000640.80-332351.40                   \\
KIDS J022735.90-315043.81                   \\
KIDS J024142.11-331023.90                   \\
KIDS J090616.66-011253.83                   \\
KIDS J091956.29+000329.12                   \\
KIDS J115341.90-021828.36                   \\
KIDS J115855.21-023108.21                   \\
KIDS J143137.93+002952.09                   \\
KIDS J151129.73-000918.63                   \\
KIDS J154257.95+003224.67                   \\
KIDS J235104.58-325322.09                   \\
KIDS J141026.79+020856.58                   \\
KIDS J085446.55-012137.14                   \\
KIDS J233105.55-335829.69                   \\
KIDS J221339.02-331156.34                   \\
KIDS J115353.49+021349.58                   \\
KIDS J140541.82+011206.58                   \\
KIDS J021035.24-321731.56                   \\ \hline
overlapped with supplementary candidates    \\
KIDS J140452.31+005122.57                   \\
KIDS J222755.52-324206.58                   \\
KIDS J021751.64-310252.07                   \\
KIDS J030557.27-342540.86                   \\
KIDS J092237.24+000313.71                   \\
KIDS J115252.26+004733.24                   \\ \hline
\end{tabular}
\caption{The overlapped sources that are found by us and previous studies  \citep{Petrillo2017,Petrillo2019}. The first 21 sources are the sources that overlap with our high probability candidates and the last 6 overlap with our supplementary candidates.}
\label{tab:over-lap}
\end{table}

\bsp
\label{lastpage}
\end{document}